\pdfoutput=1

\documentclass[11pt,twoside,a4paper,cmspaper,final,collab]{cms-tdr}
\def\svnVersion{325187}\def\cmsCernNo{2013-037}\def\cmsCernDate{\today}\def\cmsMessage{Published in Physical Review D as \href{http://dx.doi.org/10.1103/PhysRevD.93.012003}{\doi{10.1103/PhysRevD.93.012003.}}}

\begin{document}\cmsNoteHeader{B2G-13-005}

\hyphenation{had-ron-i-za-tion}
\hyphenation{cal-or-i-me-ter}
\hyphenation{de-vices}
\RCS$Revision: 325060 $
\RCS$HeadURL: svn+ssh://svn.cern.ch/reps/tdr2/papers/B2G-13-005/trunk/B2G-13-005.tex $
\RCS$Id: B2G-13-005.tex 325060 2016-02-12 00:33:10Z imarches $
\newlength\cmsFigWidth
\ifthenelse{\boolean{cms@external}}{\setlength\cmsFigWidth{0.85\columnwidth}}{\setlength\cmsFigWidth{0.4\textwidth}}

\providecommand{\cPQT}{\ensuremath{\cmsSymbolFace{T}}\xspace}
\providecommand{\cPaQT}{\ensuremath{\overline{\cmsSymbolFace{T}}}\xspace}

\providecommand{\Mfit}{\ensuremath{M_\text{fit}\xspace}}
\providecommand{\TTbar}{\ensuremath{\cPQT\cPaQT}\xspace}
\providecommand{\bW}{\ensuremath{\PQb\PW}\xspace}
\providecommand{\tZ}{\ensuremath{\PQt\cPZ}\xspace}
\providecommand{\tH}{\ensuremath{\PQt\PH}\xspace}
\newcommand{\hT}{\ensuremath{H_{\mathrm{T}}}\xspace}
\newcommand{\sT}{\ensuremath{S_{\mathrm{T}}}\xspace}
\providecommand{\FASTJET}{\textsc{fastjet}\xspace}
\newcommand{\boxCheck}{$\text{\rlap{$\checkmark$}}\square$}
\newcommand{\mgg}{\ensuremath{m_{\cPgg\cPgg}}}

\providecommand{\twothird}{\ensuremath{2\hspace{-0.06em}/\hspace{-0.08em}3}\xspace}

\ifthenelse{\boolean{cms@external}}{\providecommand{\cmsLeft}{top}}{\providecommand{\cmsLeft}{left}}
\ifthenelse{\boolean{cms@external}}{\providecommand{\cmsRight}{bottom}}{\providecommand{\cmsRight}{right}}
\ifthenelse{\boolean{cms@external}}{\providecommand{\cmsTopLeft}{top}}{\providecommand{\cmsTopLeft}{top left}}
\ifthenelse{\boolean{cms@external}}{\providecommand{\cmsTopRight}{middle}}{\providecommand{\cmsTopRight}{top right}}
\ifthenelse{\boolean{cms@external}}{\providecommand{\cmsBottom}{bottom}}{\providecommand{\cmsBottom}{bottom}}
\ifthenelse{\boolean{cms@external}}{\providecommand{\CL}{C.L.\xspace}}{\providecommand{\CL}{CL\xspace}}
\ifthenelse{\boolean{cms@external}}{\providecommand{\NA}{\ensuremath{\cdots}}\xspace}{\providecommand{\NA}{---\xspace}}
\providecommand{\CLS}{CL$_\mathrm{S}$\xspace}

\newcolumntype{.}[1]{D{.}{.}{#1}}
\newcolumntype{x}[1]{D{,}{\,}{#1}}
\newcolumntype{y}[1]{D{,}{\,\pm\,}{#1}}

\cmsNoteHeader{B2G-13-005}
\title{\texorpdfstring{Search for vector-like charge \twothird \cPQT quarks in proton-proton collisions at $\sqrt{s} = 8 \TeV$}{Search for vector-like charge 2/3 T quarks in proton-proton collisions at sqrt(s) = 8 TeV}}

\date{\today}

\abstract{
A search for fermionic top quark partners \cPQT of  charge \twothird is presented. The search is carried out in proton-proton collisions  corresponding to an integrated luminosity of  19.7\fbinv collected at a center-of-mass energy of $\sqrt{s} = 8\TeV$ with the CMS detector at the LHC. The \cPQT quarks are assumed to be produced strongly in pairs and can decay into \tH, \tZ, and \bW. The search is performed in five exclusive channels: a single-lepton channel, a multilepton channel, two all-hadronic channels optimized either for the \bW or the \tH decay, and one channel in which the Higgs boson decays into two photons. The results are found to be compatible with the standard model expectations in all the investigated final states. A statistical combination of these results is performed and lower limits on the \cPQT quark mass are set. Depending on the branching fractions, lower mass limits between 720 and 920\GeV at 95\% confidence level are found. These are among the strongest limits on vector-like \cPQT quarks obtained to date.
}

\hypersetup{%
pdfauthor={CMS Collaboration},%
pdftitle={Search for vector-like charge 2/3 T quarks in proton-proton collisions at sqrt(s) = 8 TeV},%
pdfsubject={CMS},%
pdfkeywords={CMS, physics, exotic quarks}}

\maketitle
\section{Introduction}
The discovery of a Higgs boson with a mass of 125\GeV by the ATLAS
\cite{Aad:2012tfa}  and CMS \cite{Chatrchyan:1471016,Chatrchyan:2013lba} collaborations motivates the
search for exotic states involving the newly discovered particle.  The
nature of electroweak symmetry breaking and the  mechanism that
stabilizes the mass of the Higgs particle are not entirely clear. These questions could be explained by physics
beyond the standard model (SM), such as supersymmetry. Non-supersymmetric explanations
are given by little Higgs models \cite{ArkaniHamed:2001nc,Schmaltz:2005ky}, models with extra
dimensions \cite{Antoniadis:2001cv,Hosotani:2004wv}, and composite
Higgs models
\cite{Antoniadis:2001cv,Hosotani:2004wv,Agashe:2004rs} in which the
Higgs boson appears as a pseudo-Nambu-Goldstone boson \cite{Hill:2002ap}. These theories
predict the existence of heavy
vector-like quarks. The left-handed and right-handed components of vector-like quarks transform in the same way under the electroweak symmetry group, in contrast to the SM fermions, which transform as chiral particles under the SM  symmetry group $\mathrm{ SU(3)_c \times SU(2)_L \times U(1)_Y }$.  This property of the vector-like quarks allows direct
mass terms in the Lagrangian of the form $m \overline{\psi}\psi$ that
do not violate gauge invariance. As a consequence, and in contrast to
the other quark families,  vector-like quarks do not
acquire their mass via Yukawa couplings. In many of the models mentioned above the vector-like quarks
 couple predominantly to the third generation quarks only. This means
that they  may have the following three decay modes: \tH, \tZ, and \bW~\cite{PhysRevD.88.094010}.  A model of
vector-like \cPQT quarks with charge $\twothird\, e$, which are produced in pairs via strong
interaction, is used as a benchmark for this analysis.

A fourth generation of chiral fermions, replicating one of the three generations
of the SM with identical quantum numbers, is disfavored by electroweak fits within the framework of the SM~\cite{PhysRevD.86.013011}. This is mostly because of large modifications of the Higgs production cross sections and branching fractions ($\mathcal{B}$), if a single SM-like Higgs doublet is assumed. Heavy vector-like quarks
decouple from low energy loop-level electroweak corrections and are not similarly constrained by the measurements of the Higgs boson properties~\cite{PhysRevD.88.094010}.

Early \cPQT quark searches by the CMS Collaboration \cite{Chatrchyan2012307,Chatrchyan2012103,PhysRevLett.107.271802} have assumed 100\% branching fractions to various final states. More recent searches \cite{tagkey2014149} do not make specific assumptions for the branching fractions. Searches for \cPQT quarks have been performed also by the ATLAS Collaboration, setting lower limits on the \cPQT quark mass ranging from 715 to 950\GeV, for different T quark branching fractions~\cite{Aad:2014efa,Aad:2015gdg,Aad:2015kqa}.

In this paper, results of searches for \cPQT quark production in
proton-proton collisions, using the CMS detector at the CERN LHC,  are
presented for five different decay modes. One of the searches
\cite{tagkey2014149} is inclusive and sets limits for all possible
branching fractions. This analysis is based on leptonic final states and is described in  Section~\ref{B2G12015}. The other four analyses have a good sensitivity in  optimized regions, but they do not cover the full range of branching fractions.  The analysis described in Section~\ref{B2G12017}    is specifically optimized to find $\cPQT\to \bW$ decays.  The searches presented in Section~\ref{B2G14002} and Section~\ref{B2G12013} are optimized for all-hadronic final states in the decays $ \cPQT\to \bW$ and $ \cPQT \to \tH$. The search discussed in Section~\ref{B2G14003} is sensitive to $\cPQT \to \tH$ decays, where the Higgs boson decays to a pair of photons. The two analyses presented in Sections~\ref{B2G12015} and \ref{B2G14002} are discussed in detail in separate publications~\cite{tagkey2014149,B2G14002paper}. The remaining three analysis are published here for the first time.

The CMS detector is briefly described in
Section~\ref{detector}. Section~\ref{samples} describes the data
and the simulated samples. Section~\ref{reco} gives details about the
reconstruction techniques used by the analyses.
Section~\ref{sec:combo} describes the combination and the treatment of
systematic uncertainties.  Section~\ref{sec:results} presents the
results of the combination.

\section{The CMS detector}\label{detector}
The central feature of the CMS apparatus is a superconducting solenoid of 6\unit{m} internal diameter, providing a
magnetic field of 3.8\unit{T}. Within the solenoid volume are a silicon pixel and strip tracker, a lead tungstate crystal electromagnetic calorimeter (ECAL), and a brass and scintillator hadron calorimeter (HCAL), each composed of a barrel and two endcap sections. Muons are measured in gas-ionization detectors embedded in the steel flux-return yoke outside the solenoid. Extensive forward calorimetry complements the coverage provided by the barrel and endcap detectors.

In the region of pseudorapidity $\abs{ \eta }< 1.74$~\cite{Chatrchyan:2008zzk}, the HCAL cells have widths of 0.087 in $\eta$ and 0.087 radians in azimuth ($\phi$). In the $\eta$-$\phi$ plane, and for $\abs{\eta}< 1.48$, the HCAL cells map on to $5{\times}5$ ECAL crystals arrays to form calorimeter towers projecting radially outwards from close to the nominal interaction point. At larger values of $\abs{ \eta }$, the size of the towers increases and the matching ECAL arrays contain fewer crystals. Within each tower, the energy deposits in ECAL and HCAL cells are summed to define the calorimeter tower energies, subsequently used to provide the energies and directions of hadronic jets.

The electron momentum is estimated by combining the energy measurement
in the ECAL with the momentum measurement in the tracker. The momentum
resolution for electrons with transverse momentum $\pt \approx 45\GeV$ from $\cPZ \to \Pe\Pe$ decays ranges
from 1.7\% for nonshowering electrons in the barrel region to 4.5\%
for showering electrons in the endcaps~\cite{Khachatryan:2015hwa}.
The energy resolution for photons with transverse energy $\ET \approx 60\GeV$ varies
between 1.1\% and 2.6\% in the ECAL barrel, and from 2.2\% to 5\% in the endcaps~\cite{Chatrchyan:2013dga}.

The silicon tracker measures charged particles within the pseudorapidity range $\abs{\eta}< 2.5$. It consists of 1440 silicon pixel and 15\,148 silicon strip detector modules. For nonisolated particles of $1 < \pt < 10\GeV$ and $\abs{\eta} < 1.4$, the track resolutions are typically 1.5\% in \pt and 25--90 (45--150)\unit{\mum} in the transverse (longitudinal) impact parameter \cite{TRK-11-001}.

A more detailed description of the CMS detector, together with a definition of the coordinate system used and the relevant kinematic variables, can be found in Ref.~\cite{Chatrchyan:2008zzk}.

\section{Event samples}
\label{samples}
This analysis makes use of data recorded with the CMS detector in proton-proton collisions at a center-of-mass energy of $\sqrt{s} = 8\TeV$ corresponding to an integrated luminosity of 19.5\fbinv for the analysis described in Section \ref{B2G12015}, and  19.7\fbinv for the other analyses.

Events are selected by a multi-stage trigger system. The single-lepton channels are based on single-muon and single-electron triggers. The single-muon sample is obtained by the requirement of an isolated muon candidate, with high-level trigger thresholds of  $\pt >24\GeV$ (inclusive search, Section~\ref{B2G12015}) or $\pt >40\GeV$ (single-lepton search, Section~\ref{B2G12017}). In the electron sample, a single isolated electron trigger with $\pt >27\GeV$ is required. Multilepton events are selected by requiring at least two lepton candidates, one with $\pt>17\GeV$ and the other with $\pt>8\GeV$ in the high-level trigger. The all-hadronic final states require  large hadronic activity in the detector, namely that the scalar \pt sum of reconstructed jets is larger than 750\GeV. This quantity is evaluated in the high-level trigger from jets with $\pt >40\GeV$ using calorimeter information only. For searches in the diphoton final state, two photons are required. The photon \ET thresholds in the high-level trigger are 26 (18)\GeV and 36 (22)\GeV on the leading (subleading) photon, depending on the running period.

The contributions from SM processes are generally predicted using simulated event samples. For some backgrounds, however, the simulations are not fully reliable, and control samples of data are used to determine their contribution. The background estimation for the individual channels is discussed in Section \ref{sec:analysesInComb}.

Standard Model background events are simulated using \POWHEG~v1.0~\cite{powheg1,powheg2,powheg3} for $\ttbar$ and single~\PQt
production; \MADGRAPH~5.1~\cite{MadgraphRef} for {\PW}+jets, {\cPZ}+jets, $\ttbar \PW$, and $\ttbar \cPZ$ production; and \PYTHIA~6.426 \cite{1126-6708-2006-05-026}
for $\PW\PW$, $\PW\cPZ$, $\cPZ\cPZ$, and $\ttbar \PH$ processes.

For $\PW$+jets and $\cPZ$+jets production, samples with up to four partons are generated and merged using the MLM scheme with \kt jets~\cite{Alwall:2007fs,Alwall:2008qv}. The CTEQ6M parton distribution functions (PDF) are used for \POWHEG, while for the other generators the CTEQ6L1~\cite{1126-6708-2002-07-012} PDFs are used. In all cases, \PYTHIA~6.426 \cite{1126-6708-2006-05-026} is used to simulate the hadronization and the parton showering.

The $\cPQT\cPaQT$ signal process is simulated using \MADGRAPH~5.1, allowing up to two additional hard partons.
A series of mass hypotheses between 500 and 1000\GeV are generated in steps of 100\GeV. The inclusive cross
sections for the signal samples and the $\ttbar$  samples are calculated at
next-to-next-to-leading order (NNLO) for $\Pg \Pg \to \ttbar +
X$. The fixed-order calculations are supplemented with soft-gluon
resummations having next-to-next-to-leading logarithmic
accuracy~\cite{Czakon:2013goa}.  The  $\ttbar$ cross sections are computed based on the
\textsc{Top++}~v2.0 implementation using the MSTW2008nnlo68cl
PDFs and the 5.9.0 version of
LHAPDF~\cite{Czakon:2013goa,Czakon:2011xx}. The $\ttbar$ cross section is computed to be 252.9\unit{pb}, assuming a top quark mass of 172.5\GeV. The model-independent cross sections calculated for the signal samples are
listed in Table \ref{tab:signalCrossSections}.

\begin{table}[htb]
 \centering
\topcaption{The NNLO $ \cPQT\cPaQT$ pair production cross section for different values of the \cPQT
quark mass. }
\begin{scotch}{c.{10}}
\cPQT quark mass & \multicolumn{1}{c}{Production}   \\
    $(\GeVns)$  &    \multicolumn{1}{c}{cross section (\unit{pb})}  \\
\hline
 500 &   0.59          \\
 600 &   0.17          \\
 700 &   0.059       \\
 800 &   0.021       \\
 900 &   0.0083     \\
 1000 &  0.0034    \\
\end{scotch}
\label{tab:signalCrossSections}
\end{table}

Minimum bias interactions are generated using \PYTHIA and are superimposed on the simulated events to mimic the effect of additional proton-proton collisions  within a single bunch crossing (pileup). The pileup distributions of the simulated signal and background events
match that observed in data, with an average of 21 reconstructed collisions per beam crossing.

\section{Event reconstruction \label{reco}}
Tracks are reconstructed using an iterative tracking procedure~\cite{TRK-11-001}. The primary vertices are reconstructed with a deterministic annealing method~\cite{IEEE_DetAnnealing} from all  tracks in the event that are compatible with the location of the proton-proton interaction region.
The vertex with the highest $\sum (\pt^\text{track})^2$ is defined as the
primary interaction vertex (PV),
whose position is determined from an adaptive vertex
fit~\cite{AVFitter}.

The particle-flow event reconstruction algorithm \cite{CMS-PAS-PFT-09-001,CMS-PAS-PFT-10-001}  reconstructs and identifies each individual particle, using an optimized combination of information from the various elements of the CMS detector. The energy of muons is obtained from the curvature of the corresponding track. The energy of electrons is determined from a combination of the electron momentum at the PV as determined by the tracker, the energy of the corresponding ECAL cluster, and the energy sum of all bremsstrahlung photons spatially compatible with originating from the electron track. The energy of charged hadrons is determined from a combination of their momentum measured in the tracker and the matching ECAL and HCAL energy deposits, corrected for zero suppression effects and for the response function of the calorimeters to hadronic showers. Finally, the energy of neutral hadrons is obtained from the corresponding corrected ECAL and HCAL energies.

Muon (electron) candidates are required to originate from the PV and to be isolated within $\Delta R = \sqrt{(\Delta\eta)^{2} + (\Delta\phi)^{2}} < 0.4~(0.3)$ around the lepton direction, where $\Delta\eta$ ($\Delta\phi$) indicates the difference in pseudorapidity $\eta$ ($\phi$) from the lepton direction. The degree of isolation is quantified by the ratio of the \pt sum of all additional particles reconstructed in the isolation cone to the \pt of the lepton candidate. This ratio for a muon (electron) is required to be less than 0.12 (0.10). Together with the lepton identification requirements, the isolation conditions strongly suppress backgrounds from jets containing leptons.

Photons are identified as ECAL energy clusters not linked to the extrapolation of any charged particle trajectory to the ECAL. The energy of photons is directly obtained from the ECAL measurement, corrected for zero-suppression effects. In the ECAL barrel section, an energy resolution of about 1\% is achieved for unconverted or late-converting photons in the tens of \GeV energy range. The remaining barrel photons are measured with an energy resolution of about 1.3\% up to $\abs{\eta} = 1$, rising to about 2.5\% at $\abs{\eta} = 1.4$. In the endcaps, the resolution of unconverted or late-converting photons is about 2.5\%, while all other photons have a resolution between 3 and 4\%~\cite{CMS:EGM-14-001}.

For each event, hadronic jets are reconstructed by applying the anti-\kt (AK) algorithm \cite{Cacciari:2008gp, fastjet} and/or the Cambridge--Aachen (CA) \cite{CACluster1} jet clustering algorithms to the reconstructed particles.  The AK algorithm is used with a jet size parameter of 0.5 (AK5 jets). In some analyses both algorithms are used. The algorithms are applied independently of each other to the full set of reconstructed particles. Charged particles that do not originate from the PV are removed from the jets. The momentum of each jet is defined as the vector sum of all particle momenta in the jet cluster, and is found in the simulation to be within 5\% to 10\% of the true particle-level momentum over the whole \pt spectrum and detector acceptance. Jet energy corrections are derived from the simulation, and are confirmed with measurements of the energy balance of dijet and photon+jet events~\cite{Chatrchyan:2011ds}. The jet energy resolution is typically 15\% at 10\GeV, 8\% at 100\GeV, and 4\% at 1\TeV, to be compared to about 40\%, 12\%, and 5\% obtained when the calorimeters alone are used for jet clustering.

Neutrinos escape the detector undetected and give rise to the missing transverse momentum vector, defined as the projection on the plane perpendicular to the beams of the negative vector sum of the momenta of all reconstructed particles in an event. Its magnitude is referred to as \ETmiss.

The jets contain neutral particles from pileup events. The contribution from these additional
particles is subtracted based on the average expectation of the energy deposited from pileup
in the jet area, using the methods described in Ref. \cite{Cacciari:2008gn}.

For the identification of jets resulting from fragmentation of \PQb quarks (``\PQb jets''), an algorithm is used that combines information from reconstructed tracks and from secondary vertices, both caracterized by a displacement with respect to the PV. This information is combined into a single discriminating variable and jets are tagged as \PQb jets based on its value. The algorithm is referred to as ``combined secondary vertex tagger'' and is described in Ref.~\cite{Chatrchyan:2012btv}. In most of the analyses described in the following, a minimum value of this variable (medium operating point) is chosen such that the \PQb tagging efficiency  is 70\% and the light-flavor jet misidentification rate is 1\% in \ttbar events. The analyses presented in Sections~\ref{B2G12017} and \ref{B2G14003} also use a smaller minimum value of the discriminating variable (loose operating point), yielding a higher efficiency of approximately 80\%, with a light-flavor misidentification rate of 10\%.

\subsection{Jet substructure methods}
\label{substructure}
Because of the possible large mass of the \cPQT quarks, the top quarks,  Higgs and \PW bosons from \cPQT quark decays might have  significant Lorentz boosts. Daughter particles produced in these decays would therefore not be well separated. In many cases, all decay products are clustered into a single large jet  by the event reconstruction algorithms. These merged jets exhibit  an intrinsic substructure that can be analyzed with dedicated jet substructure algorithms.  In order to cluster the decay products from top quarks and Higgs boson into wide jets, the CA algorithm is used with size parameters R=1.5 (CA15 jets) or R=0.8 (CA8 jets). A number of jet substructure algorithms are then used in different analyses to identify jets from top quark or Higgs boson decays. This process is
known as \PQt or \PH tagging, and in some cases relies on \PQb tagging of individual subjets.

The inclusive \cPQT quark search in final states with leptons discussed in Section \ref{B2G12015} uses the \textsc{CMSTopTagger}~\cite{Chatrchyan:2012ku}, which is based on the algorithm developed in Ref.~\cite{Kaplan:2008ie}. The tagger identifies a top quark decay if a CA8 jet with $\pt > 400\GeV$ is found with a mass between 140 and 250\GeV and at least three subjets with a minimum mass of  subjet pairs  larger than 50\GeV. The sensitivity of the \textsc{CMSTopTagger} is suitable for a regime with jet $\pt > 400\GeV$  where the decay products are collimated to be within the acceptance of a jet with the size parameter of 0.8.

The search for $\cPQT \to \PQt \PH$  in the hadronic final state (Section \ref{B2G14002}) adopts the \textsc{HEPTopTagger} algorithm~\cite{Plehn:2011tg, CMS-PAS-JME-13-007}, which employs CA15 jets  to increase the acceptance to top quarks with a moderate Lorentz boost ($\pt > 200\GeV$). This facilitates a smooth transition between the boosted and resolved regimes. A CA15 \PQt jet candidate is required to exhibit a substructure compatible with a three-body decay. If this requirement is satisfied, the \textsc{HEPTopTagger} clustering algorithm identifies the three subjets, and then requires that the mass of a subjet pair be consistent with the \PW boson mass and the mass of the three subjets be consistent with the top mass. The \PQt tagging performance is further enhanced by the application of \PQb tagging to subjets of CA15 jets~\cite{CMS-PAS-BTV-13-001}. Subjet \PQb tagging is also used to identify decays of boosted Higgs bosons into a bottom quark-antiquark pair. The subjets of CA15 jets are reconstructed using the filtering algorithm described in Ref.~\cite{boostedhiggs}. Two filtered subjets of CA15 jets are required to have a di-subjet invariant mass larger than 60\GeV. Both subjets are tagged using the subjet \PQb tagging algorithm, which is based on the same algorithm used for regular anti-\kt jets, discussed above, with the difference that only tracks and secondary vertices associated with the individual subjets are used to build the \PQb tag discriminator.

For the identification of boosted \PW bosons, two subjets are required to be reconstructed by a pruning algorithm~\cite{boostedhiggs,jetpruning1,jetpruning2}. The mass of the pruned jet has to be compatible with the mass of the \PW boson, within a mass window that differs slightly depending on the analysis considered. The inclusive  analysis in Section \ref{B2G12015} requires a \PW jet to have $\pt > 200\GeV$ and a mass between 60 and 130\GeV. The search for $\cPQT \to \bW$ with single leptons (Section \ref{B2G12017}) applies the same \pt selection, but the mass window is tightened to  60 to 100\GeV. The search for $\cPQT \to \bW$ in hadronic final states (Section \ref{B2G12013}) requires $\pt > 150\GeV$ in combination with a jet mass $m_\mathrm{j}$ requirement of $60 < m_\mathrm{j} < 100\GeV$. Additionally, this analysis complements pruning with a selection on the mass drop~\cite{boostedhiggs}, which is defined as the ratio of the largest subjet mass to that of the original jet. Requiring the mass drop to be $<$0.4 rejects events containing  massive jets  from  QCD multijet processes.

The different performance of the \PQt tagging and \PW tagging algorithms in data and simulation is taken into account with scale factors  that are applied to the simulated events~\cite{CMS-PAS-JME-13-007,JME13006}.

\section{Analysis channels \label{sec:analysesInComb}}

In this Section, five distinct searches for \cPQT quarks are presented, each optimized for a different topology. The analyses described in Sections~\ref{B2G12015} and \ref{B2G12017} are based on leptonic final states. While the former is an inclusive search covering all possible decay modes, the latter is a search specifically optimized to find $ \cPQT\to \bW$ decays. The searches presented in Section~\ref{B2G14002} and Section~\ref{B2G12013} are optimized for boosted event topologies in hadronic final states and make use of jet substructure techniques. Finally, the search treated in Section~\ref{B2G14003} is sensitive to $\cPQT \to \tH$ decays, where the Higgs boson decays to a pair of photons.

\subsection{Inclusive search with single and multiple
  leptons \label{B2G12015}}
The inclusive search described in this Section is sensitive to all decay modes of the \cPQT quark,  \ie, $\cPQT \to \tH$, $\cPQT \to \tZ$, and $ \cPQT \to \bW$.  It is divided into two channels: one channel in which exactly one lepton is selected and the other channel with at least two leptons. Further details are given in Ref.~\cite{tagkey2014149}.

\subsubsection{Single-lepton channel}
Single-lepton events must contain  exactly one isolated muon or electron with $\pt>32\GeV$. In addition to the lepton, events must also have at least three AK5 jets with $\pt > 120$, 90, and~50\GeV. A fourth AK5 jet with $\pt > 35\GeV$ is required if no \PW jet is identified in the event. To fulfill the lepton isolation requirement, jets must be separated by $\Delta R > 0.4$ from muons and by $\Delta R > 0.3$ from electrons. The requirement on the jet multiplicity and \pt significantly suppresses background processes. The contribution from QCD multijet events is further reduced by selecting events with $\ETmiss > 20\GeV$. The major selection requirements are summarized in Table~\ref{tab:singleselection}.

\begin{table}[htb]
\centering
\topcaption{Main selection requirements for the single-lepton analysis.}\label{tab:singleselection}
\begin{scotch}{lc}
Variable & Selection \\
\hline
\pt lepton & $>$32\GeV \\
Number of jets & $\ge$3\\
\pt jets & $>$120, 90, and 50\GeV \\
\PW tag & $\ge$1 or $\ge$1 jets with $\pt>35\GeV$\\
$\MET$ & $>$20\GeV \\
\end{scotch}                                                                                                                           \end{table}

Some background events from $\PW$+jets production remain after the event selection. This process is not well modeled by simulations and the normalization is determined from a control  sample in data. This sample is defined by single-lepton events fulfilling the signal selection criteria, but failing the requirement that a fourth jet with $\pt > 35\GeV$ or alternatively a \PW jet is identified in the event.

A boosted decision tree (BDT)~\cite{Hocker:2007ht} is used to discriminate between signal and background events.  Different BDTs are implemented for events with and without identified \PW\ jets and for each hypothetical value of the mass of the \cPQT quark. The use of dedicated BDTs for different \cPQT quark decay modes does not improve the performance, so the BDTs are trained irrespective of the branching fraction of the \cPQT quark.

The variables used for the calculation of  the BDT discriminant are jet multiplicity, \PQb-tagged jet multiplicity, \MET, lepton \pt, \pt of the third jet, \pt of the fourth jet, and \hT, where \hT is  defined as the scalar \pt sum of all jets with $\pt > 30\GeV$. For events with at least one \PW jet, the multiplicity and \pt of $\PW$-tagged  jets and the numbers of $\PQt$-tagged jets are also included in the BDT training. These variables are  chosen   based on their discrimination power  as calculated by the BDT algorithm, and on the absence of significant correlations between the different variables. The final BDT distributions are shown in Ref.~\cite{tagkey2014149}.  The total numbers of events predicted for background processes and observed in collision data are shown in Table~\ref{tab:yields12015single}. The predicted contributions for each background process are available in Ref.~\cite{tagkey2014149}. The signal selection efficiencies are between 7.5\% and 9.4\% which corresponds to an expected number of 850 events for a \cPQT quark mass of 500\GeV and 6 events for a \cPQT quark mass of 1000\GeV  assuming branching fractions to \tH, \tZ, and \bW of 25\%, 25\%, and 50\%, respectively. A detailed table with selection efficiencies and expected number of events is available in Ref.~\cite{tagkey2014149}.

\begin{table}[htb]
\centering
\topcaption{Numbers of events predicted for background processes and observed in collision data for the single-lepton analysis. The uncertainties include those in the luminosity, the cross sections and the correction factors on lepton and trigger efficiencies. From Ref.~\cite{tagkey2014149}.}\label{tab:yields12015single}
\begin{scotch}{lcc}
                & Muon & Electron \\
\hline
Total background & $61900\,\pm\,13900$        & $61500\,\pm\,13700$ \\
Data            & $58478$  &      $57743$  \\
\end{scotch}

\end{table}

\subsubsection{Multilepton channel}

This channel uses  four mutually exclusive  subsamples with at least two leptons: two opposite-sign dilepton samples (referred to as \textit{OS1} and \textit{OS2} samples) which differ by the required numbers of jets in the event, a same-sign dilepton sample (the \textit{SS} sample) and a multilepton sample. The division into opposite- and same-sign dilepton events is based on the charge of the leptons.

Multilepton events must contain at least three leptons with $\pt>20\GeV$. To reject backgrounds from heavy-flavor resonances and low-mass Drell--Yan (DY) production, multilepton events must contain a dilepton pair of the same flavor and of opposite charge with an invariant mass above 20\GeV. Events in which $\MET\leq30\GeV$ are discarded. Jets must be separated by $\Delta R > 0.3$ from the selected leptons and at least one of the jets has to fulfill the \PQb tagging criteria.

The \textit{OS1} dilepton sample targets events in which both \cPQT quarks decay to \bW~\cite{Chatrchyan2012103}.  This dilepton sample contains events with either two or three jets, $\hT>300\GeV$, and $\sT>900\GeV$, where \sT is the sum of \hT, \MET, and the  transverse momenta of all leptons. Events are discarded where there is a dilepton pair with same-flavor leptons and a mass $M_{\ell\ell}$ consistent with that of a \cPZ boson ($76 < M_{\ell\ell} < 106\GeV$). To reduce the \ttbar  background, all the possible pair-wise combinations of a lepton and a \PQb jet are considered and their invariant masses are all required to be larger than 170\GeV.

The DY background is not modeled reliably in the selected kinematic region and is controlled using a data sample consisting of events with no $\PQb$-tagged jets, $\MET<10\GeV$, $\sT < 700\GeV$, and $\hT > 300\GeV$.

The \textit{OS2} dilepton sample consists of events with at least five jets, two of which must be  identified as \PQb jets. Events are also required to have $\hT>500\GeV$, and $\sT>1000\GeV$. This sample is mostly sensitive to signal events where both \cPQT quarks decay to \tZ. The dominant background is  $\ttbar$ production.

The \textit{SS} sample selection criteria target events in which at least one \cPQT quark decays to \tZ or \tH. Besides the lepton selection criteria, at least three jets are required, $\hT>500\GeV$, and $\sT>700\GeV$.

Different processes contribute to the background in the \textit{SS} sample. A minor contribution is given by SM processes leading to prompt \textit{SS} dilepton signatures, which have very small cross sections. These processes can be simulated reliably. The prompt \textit{OS} dilepton production can also contribute if one lepton is misreconstructed with the wrong sign of the charge. The misreconstruction probability of the charge sign is negligible for muons in the kinematic range considered, while for electrons it is determined from control data samples. We determine the probability to misreconstruct the charge sign of an electron from events with a dileptonic \cPZ decay, selected with the same criteria as in the signal selection except for the charge requirement. Instrumental backgrounds in which misidentified jets create lepton candidates are determined from control data samples in which non-prompt and fake leptons are enriched.

The multilepton sample, like the \textit{SS} sample, is mostly sensitive to signal events in which at least one \cPQT quark decays to \tZ or \tH. The backgrounds are suppressed by selecting events with at least three jets,  $\hT>500\GeV$, and $\sT>700\GeV$. Prompt backgrounds in this channel are due to SM processes with three or more leptons in the final state, such as diboson and triboson production. These are correctly modeled  by simulation. Nonprompt backgrounds are caused by the misidentification of one or more leptons, by \ttbar production, and by other processes. As for the dilepton samples, data control samples are used to evaluate these sources of background.

The main selection requirements for the four samples are summarized in Table~\ref{tab:multiselection}.

\begin{table*}[htb]
\centering
\topcaption{Main selection requirements for the four multilepton channels: the opposite-sign dilepton samples with two or three jets (\textit{OS1}) and with at least five jets (\textit{OS2}), the same-sign dilepton sample (\textit{SS}), and the multilepton sample. The smallest mass obtained  from all the possible combinations of leptons and \PQb jets is indicated by $M_{\PQb\ell}$.}\label{tab:multiselection}

\begin{scotch}{lrrrr}
          & \multicolumn{1}{c}{\textit{OS1}} & \multicolumn{1}{c}{\textit{OS2}} & \multicolumn{1}{c}\textit{SS} & \multicolumn{1}{c}{Multileptons} \\
\hline
$\hT$ (\GeVns) & $>$300 & $>$500 & $>$500 & $>$500 \\
$\sT$ (\GeVns) & $>$900 & $>$1000 & $>$700 & $>$700 \\
Number of jets & \multicolumn{1}{l}{2 or 3} & $\ge$5 & $\ge$3 & $\ge$3 \\
\PQb tags & $\ge$1 & $\ge$2 & $\ge$1 & $\ge$1 \\
\MET (\GeVns)&  $>$30 &  $>$30 &  $>$30 &  $>$30 \\
$M_{\PQb\ell}$ (\GeVns) & $>$170 & \NA & \NA & \NA \\
$M_{\ell\ell}$ (\GeVns) & $>$20 & $>$20 & $>$20 & $>$20 \\
\cPZ{} boson veto &yes & no & no & no \\
\end{scotch}

\end{table*}

The numbers of events in the multilepton samples are given in Table~\ref{tab:yields_ll}, both for data and for estimated background contributions. The predicted contributions for each background process are available in Ref.~\cite{tagkey2014149}.  The selection efficiencies for signal events are between 0.15\% and 0.44\% which corresponds to an expected number of 16.7 events for a \cPQT quark mass of 500\GeV and 0.28 events for a \cPQT quark mass of 1000\GeV, assuming branching fractions to \tH, \tZ, and \bW of 25\%, 25\%, and 50\%, respectively. A detailed table with selection efficiencies and expected number of events is available in Ref.~\cite{tagkey2014149}. The numbers of background and signal events are of similar order of magnitude.  The sensitivity to the signal is enhanced by further splitting the samples according to the lepton flavor. The dilepton samples are separated into three subsamples, $\mu\mu$, $\mu\Pe$, and $\Pe\Pe$. The multilepton sample is divided into a $\mu\mu\mu$ subsample, an $\Pe\Pe\Pe$ subsample, and a third subsample with events with mixed lepton flavors.  Data and SM background expectations are found to be in agreement.

\begin{table*}[htb]
\centering
\topcaption{Numbers of events selected in data and expected for the backgrounds. Shown are the opposite-sign dilepton samples with two or three jets (\textit{OS1}) and with at least 5 jets (\textit{OS2}), the same-sign dilepton sample (\textit{SS}), and the multilepton sample. The background sources not contributing to the channel are indicated by a dash (``--''). The uncertainties include statistical, normalization, and luminosity uncertainties. From Ref.~\cite{tagkey2014149}.}\label{tab:yields_ll}

\begin{scotch}{lcccc}
                       & \textit{OS1} & \textit{OS2} & \textit{SS}     & Multileptons \\
\hline
Total background         & $17.4\,\pm\,3.7$     & $84\,\pm\,12$  & $16.5\,\pm\,4.8$              & $3.7\,\pm\,1.3$ \\
Data                    &  20 & 86 & 18 & 2 \\
\end{scotch}

\end{table*}

\subsection{Search for \texorpdfstring{$ \cPQT\to \bW $}{T to bW} with single leptons \label{B2G12017}}
The analysis  described in this Section is optimized for the event topology in which both \cPQT quarks decay into a bottom quark and a \PW boson.

Events are required to  have one isolated  muon or electron, where muon
candidates must have $\pt > 45\GeV$ and electron candidates must have $\pt > 30\GeV$.  At
least four jets are required, either at least four AK5 jets or at least three AK5 jets plus
at least one CA8 jet. The AK5 jets are required to have $\pt > 30\GeV$
and CA8 jets are required to have $\pt > 200\GeV$. Both types of jets must
have $\abs{\eta} < 2.4$.

The CA8 jets are used to identify merged hadronic decays of \PW bosons with high Lorentz boost. The AK5 jets are replaced by the two pruned subjets of $\PW$-tagged CA8 jets if the angular distance between AK5 and CA8 jets fulfills the matching criterion  $\Delta R\mathrm{(Jet_{CA8},\, Jet_{AK5})} < 0.04$. Unmatched AK5 jets
and the subjets of matched $\PW$-tagged CA8 jets are used
as input for a kinematic fit, which is described below. The four jets or subjets are
required to satisfy $\pt > 120$, 90, 50, and 30\GeV.
At least one of the AK5 jets has to satisfy the \PQb tagging criteria.

A kinematic fit is made to each event for the
hypothesis $\TTbar \to \PQb\PWp \PAQb\PWm \to
\ell\nu\PQb\PQq\PAQq'\PAQb$, subject to the constraints,
$m(\ell\Pgn)= m(\PQq\PAQq')= M_{\PW}$, and $m(\ell \Pgn \PQb) = m(\PQq\PAQq'\PQb) = \Mfit$, the fitted mass of the selected \cPQT candidate.  The  \ETmiss in the event is attributed to the
undetected neutrino from leptonic \PW decays. If a selected event has more
than four jets, the fifth jet with highest \pt is also considered and all
the possible combinations of four jets are tested in the kinematic fit.

Only events containing fit combinations with $\chi^2$ probability $p(\chi^2) > 1\%$  are retained. The efficiency of the $p(\chi^2)$ criterion
 is 62\%  for signal events with a \cPQT quark mass of 800\GeV while 76\% of background events are rejected. The
$p(\chi^2)$ criterion removes badly reconstructed events with poor mass resolution
and improves the signal-to-background ratio in the reconstructed mass spectrum.

To reduce the large combinatorial background, the \PQb tagging and the \PW tagging
information is used. If a \PW tag is present, only
those combinations where the subjets of the \PW jet match
the \PW decay products are considered. The best combination is selected from
groups of fit combinations with decreasing \PQb tag multiplicity, ranked by
the \PQb tagging operating point (OP), as listed below:
\begin{itemize}
\item 2 \PQb tags at medium OP;
\item 1 \PQb tag at medium OP and 1 \PQb tag at loose OP;
\item 1 \PQb tag at medium OP;
\item 2 \PQb tags at loose OP.
\end{itemize}

\begin{figure*}[htb]
\centering
\includegraphics[width=0.495\textwidth]{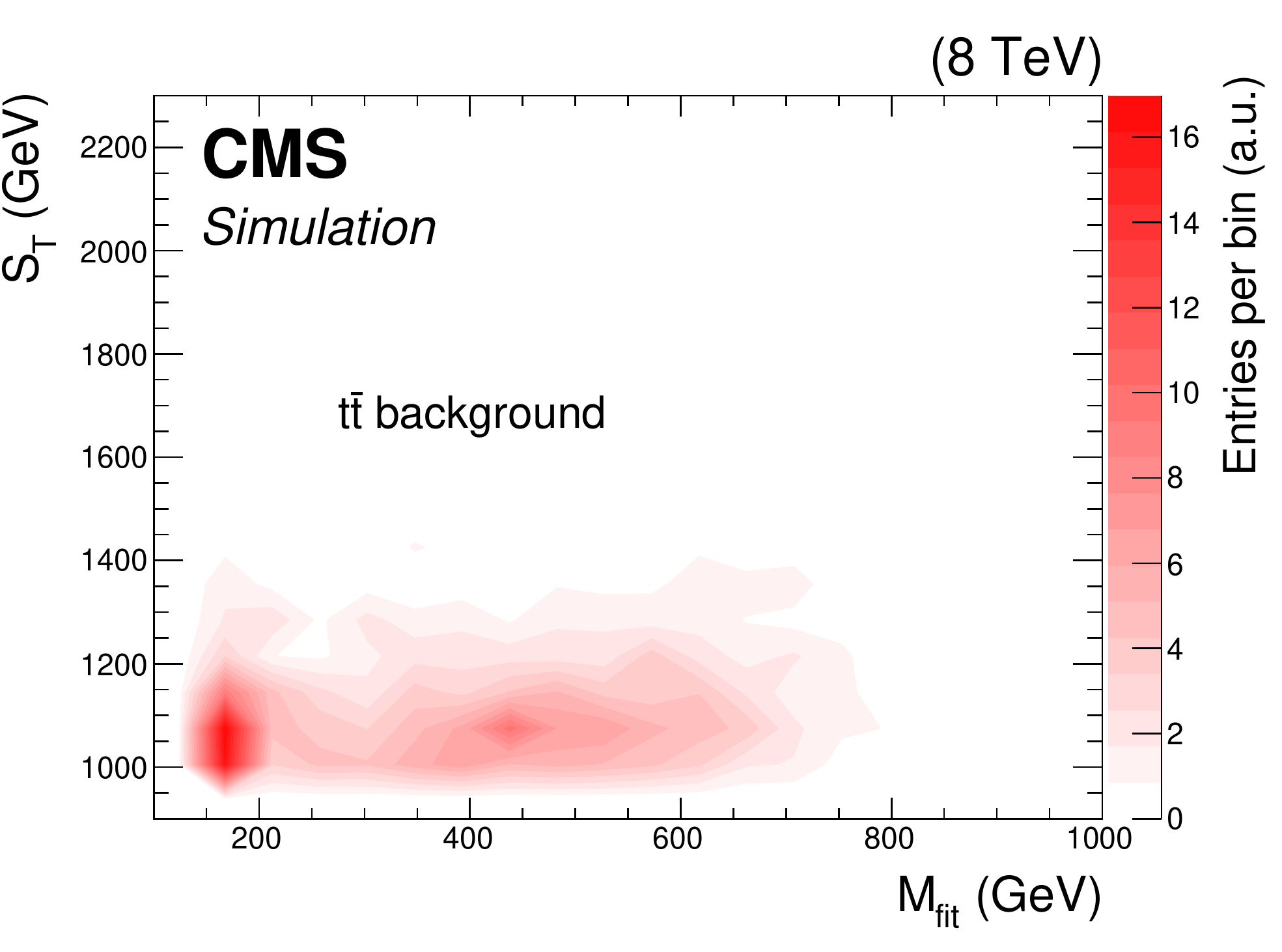}
\includegraphics[width=0.495\textwidth]{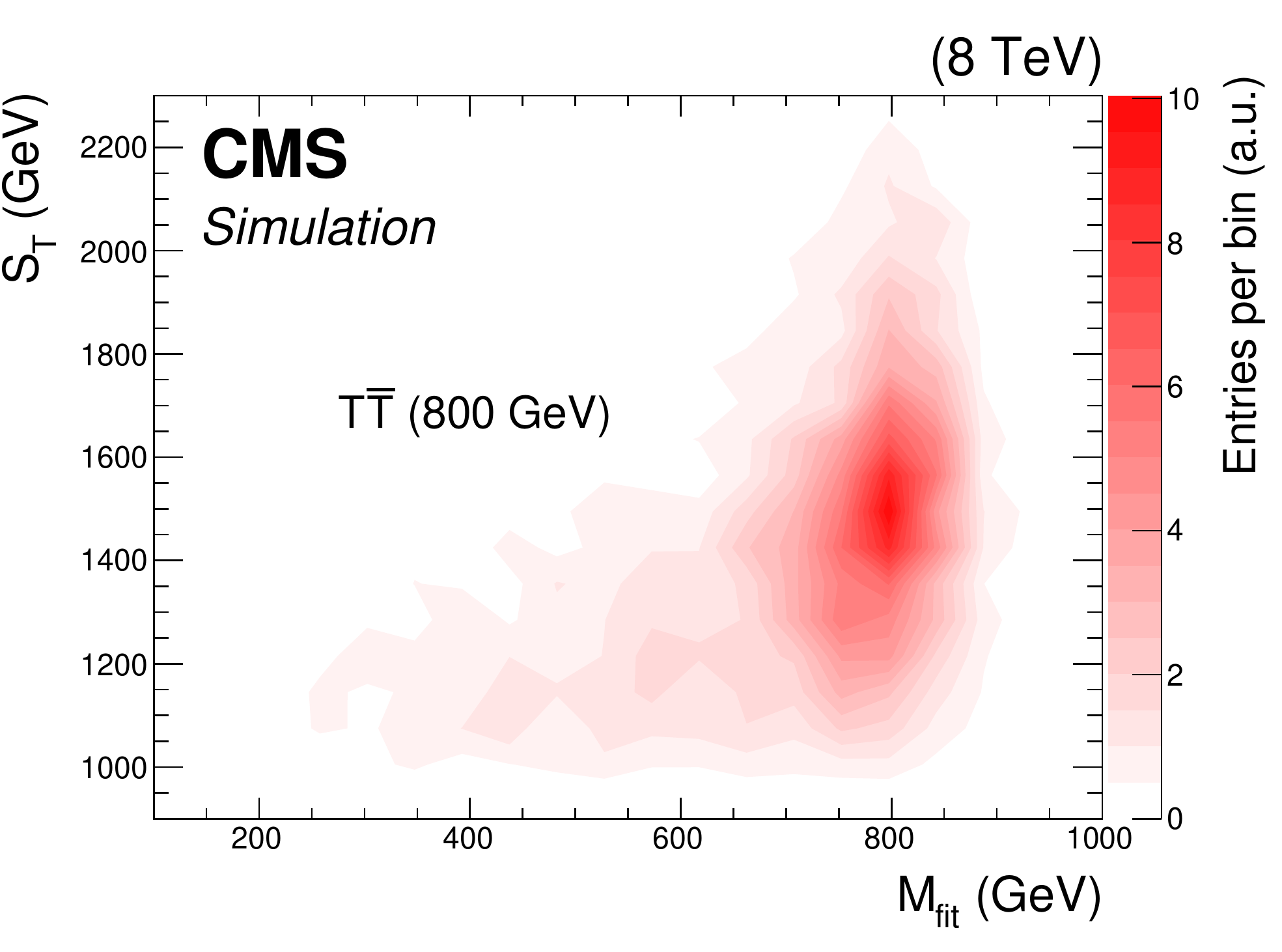}
\caption{ Correlation between the \sT and the $\Mfit$ observables in the search
for $ \cPQT\to \bW$ with single leptons, for background processes (left) and for
a simulated signal, with a \cPQT quark mass of 800\GeV (right). The color gradient indicates the entries per bin in arbitrary units (a.u.).}
\label{fig:2d_signal_bkg}

\end{figure*}

Decay products of \cPQT quarks have on average higher \pt than
those from the SM backgrounds.
To suppress the backgrounds and enhance the signal significance, we select events with large values of
the \sT variable, which is defined here as a sum of  \MET, \pt of the lepton,
and \pt of the four jets that minimize the $\chi^2$ in the kinematic fit.
Figure~\ref{fig:2d_signal_bkg}
demonstrates that SM backgrounds and a \cPQT quark signal
populate different regions in the two-dimensional \sT and $\Mfit$ distribution.

\begin{figure*}[htb]
\centering
\includegraphics[width=0.495\textwidth]{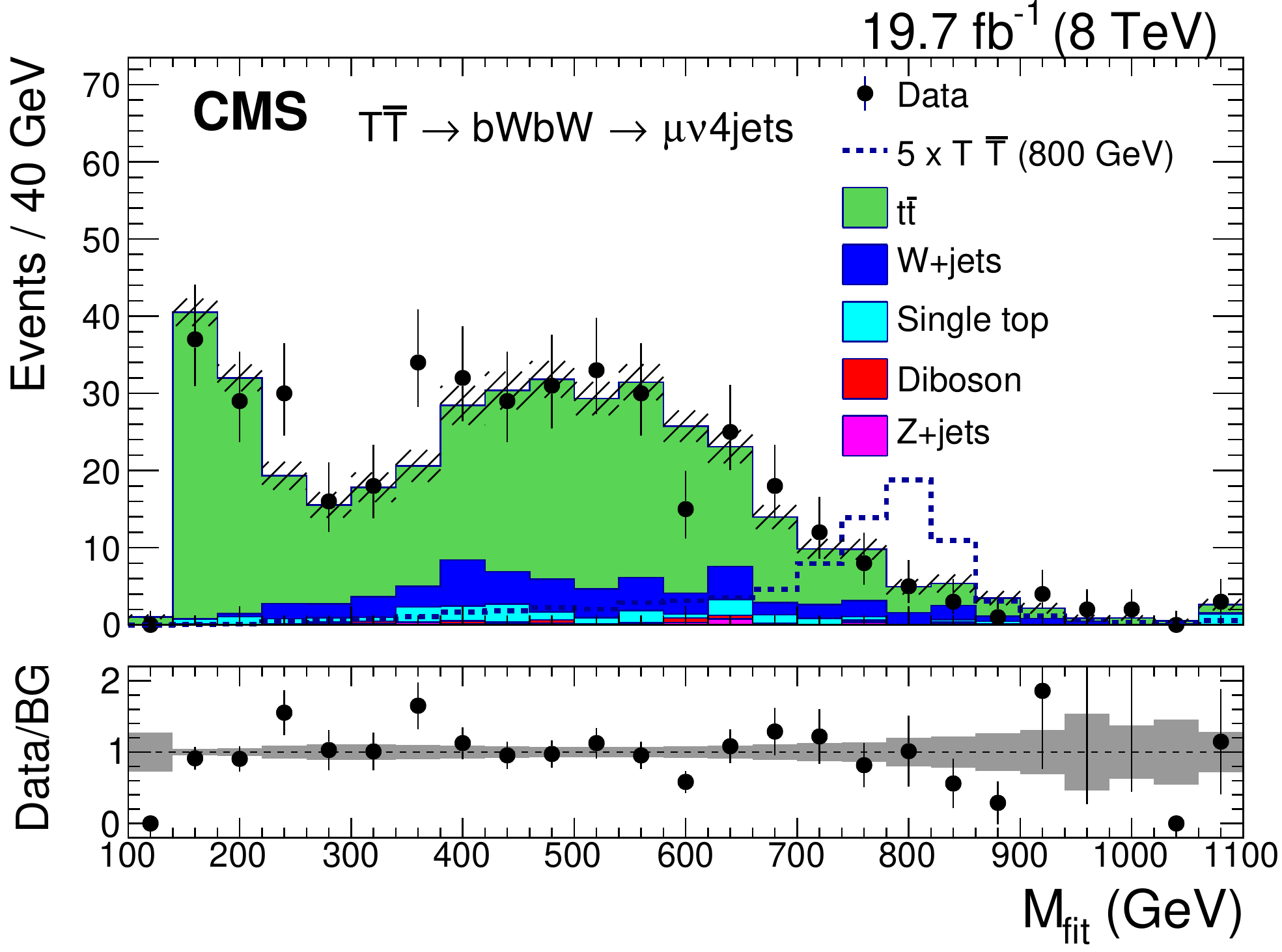}
\includegraphics[width=0.495\textwidth]{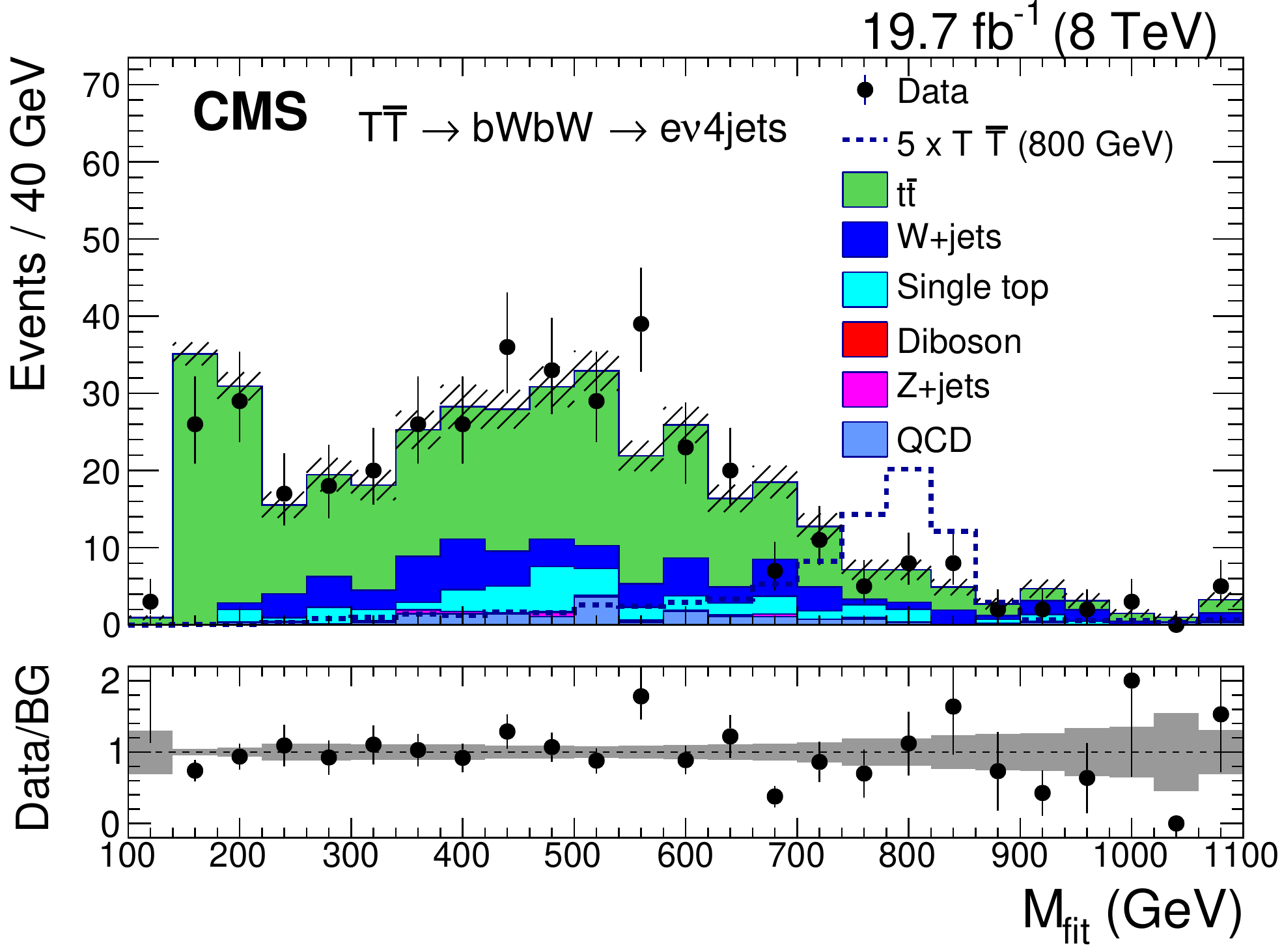}
\caption{Distributions of the reconstructed mass $\Mfit$ for $\mu$+jets
(left) and $\Pe$+jets (right) events. The data are shown as points and the simulated
backgrounds as shaded histograms. The hatched region and the shaded area in the lower panel represent the statistical uncertainty in the background.  The expected signal (dotted line) for a $\cPQT$ quark with a mass of 800\GeV is multiplied by a factor of 5 for better visibility. The lower panel represents the ratio between data and the sum of the backgrounds (BG). The overflow of the distributions is added to the last bin.}
\label{fig:mfit1}

\end{figure*}

We test the modelling of the shape of the reconstructed mass,
and verify how well the SM background expectations agree with data, as a function of \sT.
Figure~\ref{fig:mfit1} shows the reconstructed mass distributions separately for $\mu$+jets
and e+jets events  with the $\sT > 1000\GeV$ requirement.
Correctly reconstructed \ttbar events peak near the top quark mass value, while
events with mis-assigned jets constitute a combinatorial background, and
populate a region of higher masses, where the potential signal is expected to appear.
Table~\ref{tab:events} (left columns) presents the event yields of SM backgrounds and data
for this selection.
The dominant background process is $\ttbar$ production.
Smaller but still significant backgrounds come from $\PW$+jets and
single top quark production. In the $\Pe$+jets channel there is also a contribution from QCD multijet production.
Other backgrounds have been
found to be negligible.
Data and SM background expectations agree in both shape
and total normalization.

\begin{table*}[htb]
\topcaption{
Numbers of observed and expected  background events after the event selection.
The uncertainties in the  predicted numbers of events include both the statistical
and systematic uncertainties.
}
\label{tab:events}
\centering
\begin{scotch}{ly{-1}y{-1}y{-1}y{-1}}
         & \multicolumn{2}{c}{Selection ($\sT >1000\GeV$)}
& \multicolumn{2}{c}{Selection ($\sT >1240\GeV$)} \\
         & \multicolumn{1}{c}{$\mu$+jets} & \multicolumn{1}{c}{$\Pe$+jets} & \multicolumn{1}{c}{$\mu$+jets} & \multicolumn{1}{c}{$\Pe$+jets} \\
\hline
$\ttbar$    & 325,37 & 279,35  & 51,6 & 52,6 \\
$\PW + {\ge}3\text{jets}$     &  49,8  & 60,9  & 18,3  & 19,4 \\
Single top   & 20,5  & 36,10  & 6.9,2.3 & 10,4 \\
Z/$\cPgg^\ast + {\ge}3$jets & 3.9,0.8  & 3.3,0.6 & 1.4,0.4  & 1.1,0.3 \\
$\PW\PW$, $\PW\cPZ$, $\cPZ\cPZ$ & 3.1,1.0 & \multicolumn{1}{c}{$<$1}   & \multicolumn{1}{c}{$<$1} & \multicolumn{1}{c}{$<$1} \\
Multijet            & \multicolumn{1}{c}{$<$1}  & 18,4  & \multicolumn{1}{c}{$<$1}  & 6.1,1.7 \\
Total background    & 401,38 & 396,38  & 77,7 & 88,9 \\
Data       &  \multicolumn{1}{c}{417}     & \multicolumn{1}{c}{398}   & \multicolumn{1}{c}{81}      & \multicolumn{1}{c}{83} \\
\end{scotch}

\end{table*}

\begin{table}[htb]                                                                                                                                                                 
\centering                                                                                                                                                                         
\topcaption{Main selection requirements for the  $ \cPQT\to \bW$ search                                                                                                            
with single leptons.}\label{tab:singleselectionbW}                                                                                                                                 
\begin{scotch}{lc}                                                                                                                                                                 
Variable & Selection \\                                                                                                                                                            
\hline                                                                                                                                                                             
\pt muon & $>$45\GeV \\                                                                                                                                                            
\pt electron & $>$30\GeV \\                                                                                                                                                        
Number of jets & $\ge$4 \\                                                                                                                                                         
\pt jets & $>$120, 90, 50, and 30\GeV \\                                                                                                                                           
\PW tags & 0 or 1\\                                                                                                                                                                
\PQb tags & 1 or 2\\                                                                                                                                                               
\sT & $>$1240\GeV \\                                                                                                                                                               
$\MET$ & $>$30\GeV\\                                                                                                                                                               
\end{scotch}                                                                                                                                                                       
                                                                                                                                                                                   
\end{table}

\begin{table*}[htb]
\topcaption{Selection efficiencies and numbers of expected signal events
for the selection $\sT >1240\GeV$, for the two channels of the $ \cPQT\to \bW$ search
with single leptons. Different \cPQT quark  mass hypotheses  are considered and
a 100\% branching fraction to \bW is assumed.}
\label{tab:signal12017}
\centering
\begin{scotch}{ccccc}
\cPQT quark mass    &\multicolumn{2}{c}{Muon channel} &\multicolumn{2}{c}{Electron channel}\\
$(\GeVns)$ & Efficiency & Events & Efficiency & Events \\
\hline
500  &  $0.50\%$ & 59 &  $0.46\%$ & 53\\
600  &  $1.24\%$ & 43 &  $1.30\%$ & 44\\
700  &  $2.38\%$ & 28 &  $2.38\%$ & 27\\
800  &  $3.04\%$ & 13 &  $3.17\%$ & 13\\
900  &  $3.48\%$ & 5.6 & $3.63\%$ & 5.8\\
1000 &  $3.52\%$ & 2.3 & $3.86\%$ & 2.5\\
\end{scotch}

\end{table*}

We apply a requirement of $\sT >1240\GeV$ in the final event selection. This condition
is optimized to enhance the sensitivity to the signal, based on SM backgrounds and \cPQT
signal expectations.  The major selection requirements are summarized in Table~\ref{tab:singleselectionbW}.

Table~\ref{tab:events} (right columns) presents the event yields
for expected SM backgrounds and data.
Signal efficiencies
are of the order of 0.5--4\% for \cPQT quark masses from 500 to 1000\GeV.
They are summarized in Table~\ref{tab:signal12017}.

\begin{figure}[htb]
\centering
\includegraphics[width=0.49\textwidth]{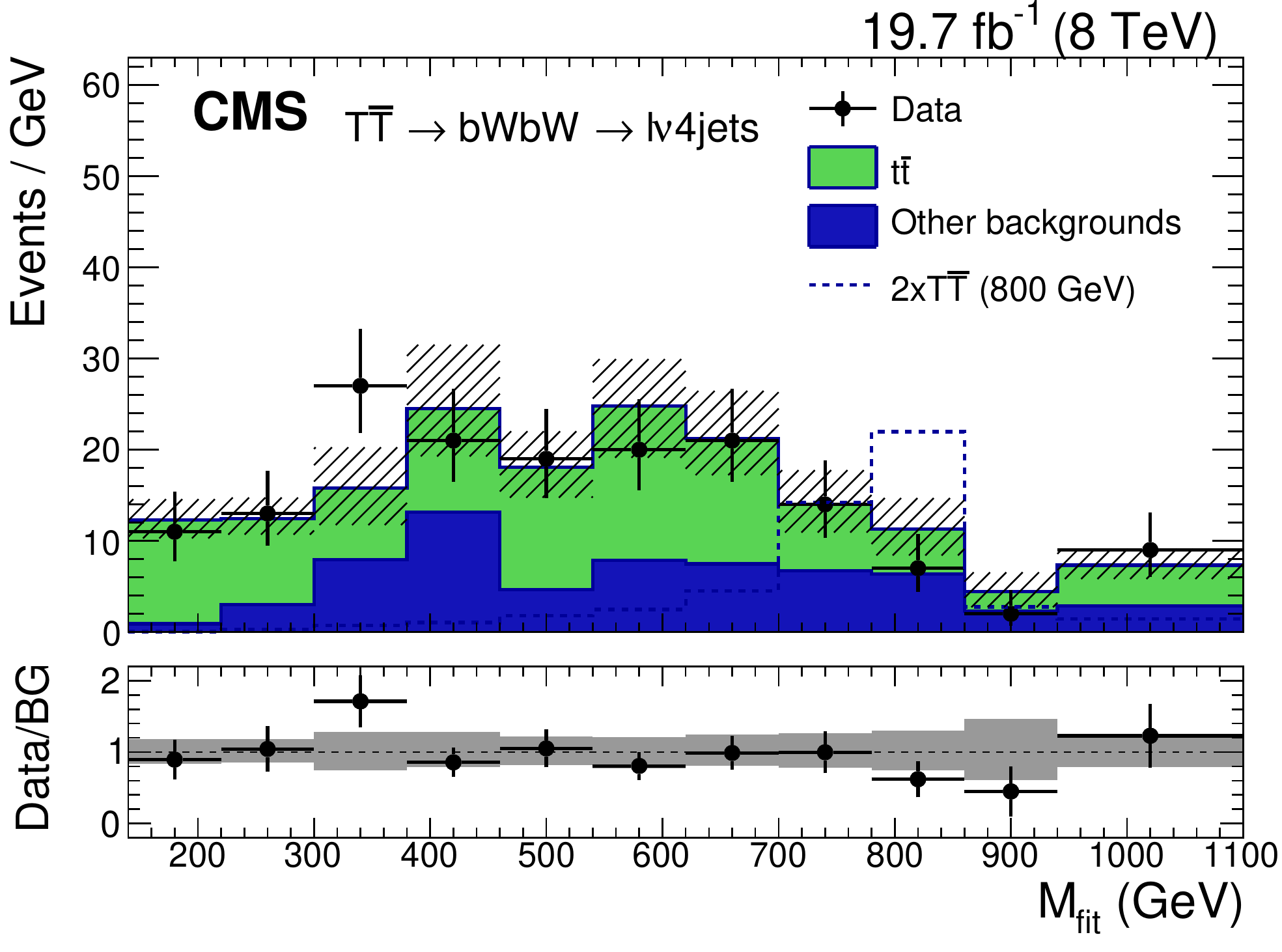}
\caption{Distributions of the reconstructed \cPQT quark mass $\Mfit$ for $\bW\bW$
candidate events in the search for $ \cPQT\to \bW$ with single leptons, combining
the $\mu$+jets and $\Pe$+jets samples after the selection $\sT > 1240\GeV$.
Data are shown as points and the simulated backgrounds as shaded histograms. The hatched region and the shaded area in the lower panel represent both the statistical and the systematic uncertainties in the total background. The expected signal for a \cPQT quark of mass 800\GeV is
multiplied by a factor of 2. The lower panel represents the ratio between data and the sum of the backgrounds (BG). The horizontal error bars represent the bin width. The overflow of the distribution is added to the last bin.}
\label{fig:mfit2}

\end{figure}

The $\Mfit$ distribution for the final event selection
is shown in Fig.~\ref{fig:mfit2}.
The $\mu$+jets and $\Pe$+jets final states give very similar results. The observed data are compatible with background expectations from SM processes.  The $\mu$+jets and $\Pe$+jets channels  are combined to improve
the statistics for the simulated SM backgrounds.

\subsection{All-hadronic search for \texorpdfstring{$ \cPQT \to \tH $}{T to tH} \label{B2G14002}}
This channel is optimized for the event topology in which at least one \cPQT quark decays to $\cPQT \to \tH$, where the top
quark decays into \bW and the \PW boson decays hadronically, and the Higgs
boson decays into two \PQb quarks. Because of the
expected high mass of the \cPQT quarks, the top quarks and Higgs bosons can have significant Lorentz boost; therefore the event selection is based on jet substructure requirements, as described in Section \ref{substructure}.

At least one $\PQt$-tagged and one $\PH$-tagged CA15 jet are required, where the $\PQt$-tagged jets must have $\pt > 200\GeV$ and the $\PH$-tagged
jets must have  $\pt > 150\GeV$.
Two variables are used to further
distinguish the signal from the background events after the event
selection. These variables are $\HT^\text{sub}$, defined here as the scalar \pt sum of subjets of  CA15 jets, and the invariant mass $m_{\bbbar}$ of two
$\PQb$-tagged subjets in the $\PH$-tagged jets. These two variables are used  for setting upper limits on the \cPQT quark production cross section.  The major selection requirements are summarized in Table~\ref{tab:14002selection}.

\begin{table}[htb]
\centering
\topcaption{Main selection requirements for the all-hadronic search for $ \cPQT\to \tH$.}\label{tab:14002selection}
\begin{scotch}{lc}
Variable & Selection \\
\hline
$\HT^\text{sub}$ & $>$720\GeV \\
Number of CA15 jets & $\ge$2 \\
\pt CA15 jets & $>$150\GeV \\
\pt $\PQt$-tagged jets & $>$200\GeV \\
Number of \PQt tags & $\ge$1 \\
Number of \PH tags & $\ge$1 \\
\end{scotch}

\end{table}

Backgrounds due to QCD multijet production are determined from data
using  signal-depleted sideband regions. These sidebands
are defined by inverting the jet substructure criteria. Backgrounds
due to \ttbar events are determined from simulation;
other backgrounds are found to be negligible.

To maximize the sensitivity of the analysis, the events are divided into two categories: a category with a single \PH tag and a category
with at least two \PH tags. The background estimates are well matched to the observed data, as discussed in Ref.~\cite{B2G14002paper}.
For the final event selection, the $\HT^\text{sub}$ and $m_{\bbbar}$ variables are combined into a single discriminator using a
likelihood ratio method. The numbers of expected background events and events observed in data after the full selection are shown in Table~\ref{tab:BackgroundsResults}. The observed data are compatible with background expectations from SM processes. The signal selection efficiencies are between 2.5\% and 7.2\% which corresponds to an expected number of 283 signal events for a \cPQT quark mass of 500\GeV and 4.9 events for a \cPQT quark mass of 1000\GeV, assuming $\mathcal{B} (\cPQT \to \tH) = 100\%$. A detailed table with selection efficiencies and expected numbers of signal events is available in  Ref.~\cite{B2G14002paper}.

\begin{table*}[htb]
 \centering
\topcaption{Predicted numbers of total background events and observed events for  the  two event categories with one and with multiple \PH tags, for the all-hadronic search for $ \cPQT\to \tH$.  The quoted uncertainties are statistical only. From Ref.~\cite{B2G14002paper}.}
\begin{scotch}{lcc}
 & Single \PH tag category & Multiple \PH tags category \\
\hline
Total background &$1403\,\pm\,14$ &  $182\,\pm\,5$  \\
Data  & 1355  &  205 \\
\end{scotch}
\label{tab:BackgroundsResults}
\end{table*}

\subsection{All-hadronic search for \texorpdfstring{$ \cPQT\to \bW$}{T to bW}}
\label{B2G12013}
This channel is optimized for the event topology in which both \cPQT quarks decay to $ \cPQT\to \bW$, where the W
bosons decay hadronically.  Events are selected by requiring two $\PW$-tagged CA8 jets with
$\pt > 150\GeV$. At least two
additional AK5 jets with $\pt > 50\GeV$ are required, one of
which must be $\PQb$-tagged. Events
are divided into categories defined by the numbers of $\PQb$-tagged jets:
one or at least two.

\begin{figure}[htb]
\centering
\includegraphics[width=0.49\textwidth]{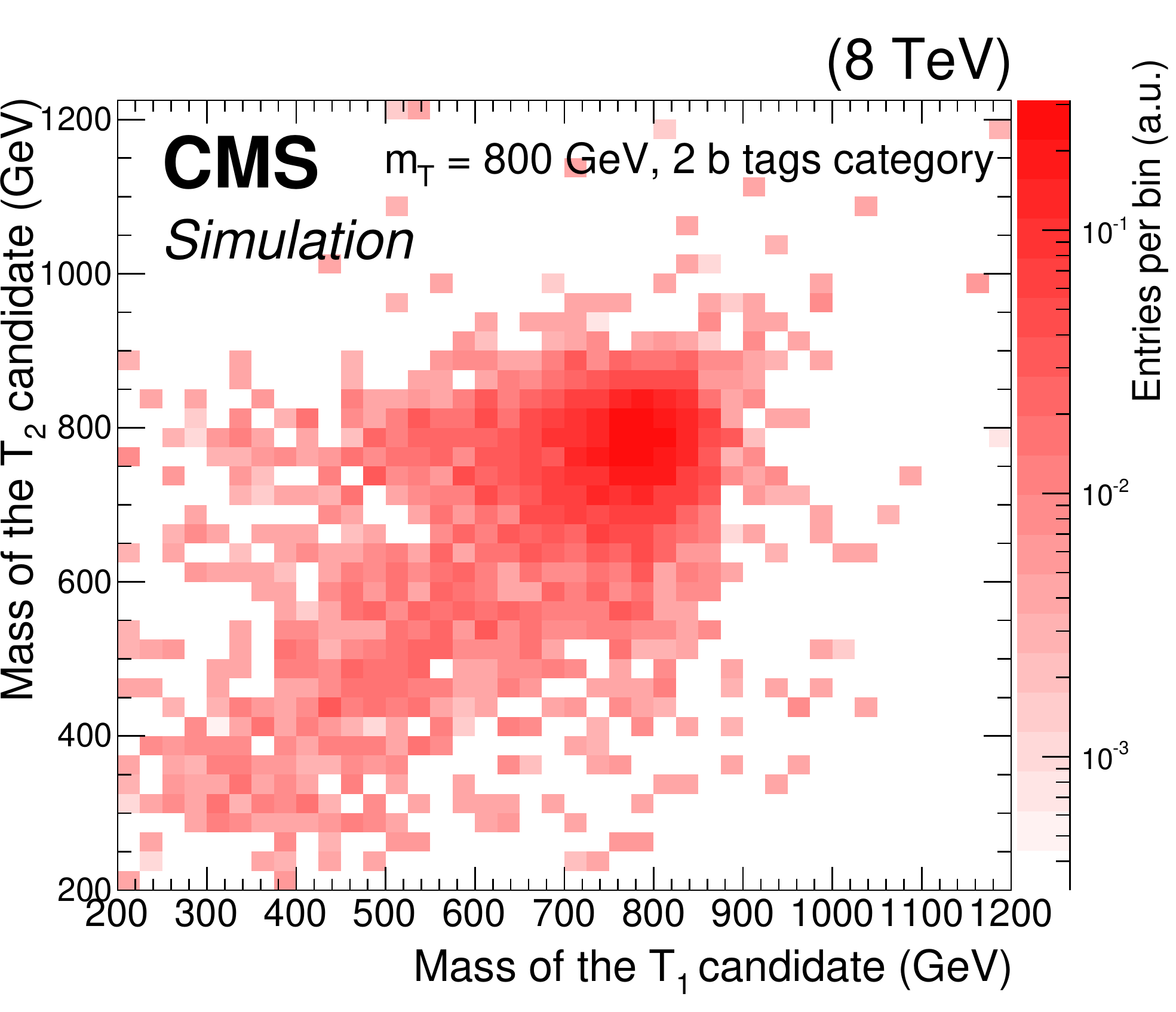}
\caption{Two-dimensional distribution of the masses of each reconstructed \cPQT candidate in the selected events for the all-hadronic search for $ \cPQT\to \bW$, for a simulated signal sample with a \cPQT quark mass of 800\GeV.  The order of  $ \cPQT_1$ and $ \cPQT_2$ is arbitrary. }
\label{fig:mWbReco_allhad}

\end{figure}

After the event selection, two \cPQT candidates $ \cPQT_1$ and $ \cPQT_2$ are
reconstructed using combinations of the \PW jets and the AK5 jets. The order of  $ \cPQT_1$ and $ \cPQT_2$ is arbitrary. The
reconstruction is performed by identifying the combinations of \PW jets
and AK5 jets having the smallest invariant mass difference.  Figure \ref{fig:mWbReco_allhad} shows the two-dimensional distribution of the masses of each reconstructed \cPQT candidate in a signal sample with a simulated \cPQT quark mass of 800\GeV.  The
reconstructed mass peak is clearly visible at the expected value. The misreconstruction
rate, where the wrong combination of jets is chosen, is small and does
not affect the signal acceptance. Additional event requirements are then
applied to increase sensitivity to the signal process.  The \cPQT
candidate masses must be greater than 200\GeV, and the fractional
difference $a_f$ in the masses of the two \cPQT candidates $m({\cPQT_1})$ and $m({\cPQT_2})$, where $a_f = |m({\cPQT_1})-m({\cPQT_2})| / (m({\cPQT_1})+m({\cPQT_2}))$, must be less than 10\%.  The two \cPQT candidates must fall in opposite hemispheres of the detector,
$\Delta\phi({\cPQT_1}, {\cPQT_2}) > 5\pi/6$, and finally $\HT^\text{4 jet}$ must be above 1000\GeV, where  $\HT^\text{4 jet}$ is defined as the scalar \pt sum of the four jets used to reconstruct the \cPQT candidates. The major selection requirements are summarized in Table~\ref{tab:12013selection}.

\begin{table}[htb]
\centering
\topcaption{Main selection requirements for the all-hadronic search for $ \cPQT\to \bW$.}\label{tab:12013selection}
\begin{scotch}{lc}
Variable & Selection \\
\hline
Number of AK5 jets & $\ge$2 \\
\pt AK5 jets & $>$50\GeV \\
Number of $\PW$-tagged jets & $\ge$2 \\
\pt $\PW$-tagged jets & $>$150\GeV \\
Reconstructed \cPQT candidate mass & $>$200\GeV \\
$a_f$ & $<$10\% \\
$\Delta\phi({\cPQT_1}, {\cPQT_2})$& ${>}5\pi/6$ \\
$\HT^\text{4 jet}$ & $>$1000\GeV \\
\end{scotch}

\end{table}

The dominant backgrounds are due to QCD multijet production and  \ttbar production. Other background contributions are negligible.

To obtain the shape of the QCD multijet background, a control region is defined by requiring $\HT^\text{4 jet} > 1000\GeV$, but inverting the requirement on the fractional mass difference, $a_f > 0.1$.  This control region is enriched in multijet events and has a negligible signal contamination.  The shape of the $\HT^\text{4 jet}$ distribution in the control region, after subtracting the expected \ttbar contribution, is used to model the QCD multijet events entering the signal region.  The $\HT^\text{4 jet}$ distribution in the signal region agrees with the distribution in the sideband region for simulated QCD multijet events.  The normalization of the QCD multijet background is not fixed, and is determined   in the limit setting procedure.  This procedure is done independently for events containing one and at least two $\PQb$-tagged jets.

\begin{figure*}[h]
\centering
\includegraphics[width=0.495\textwidth]{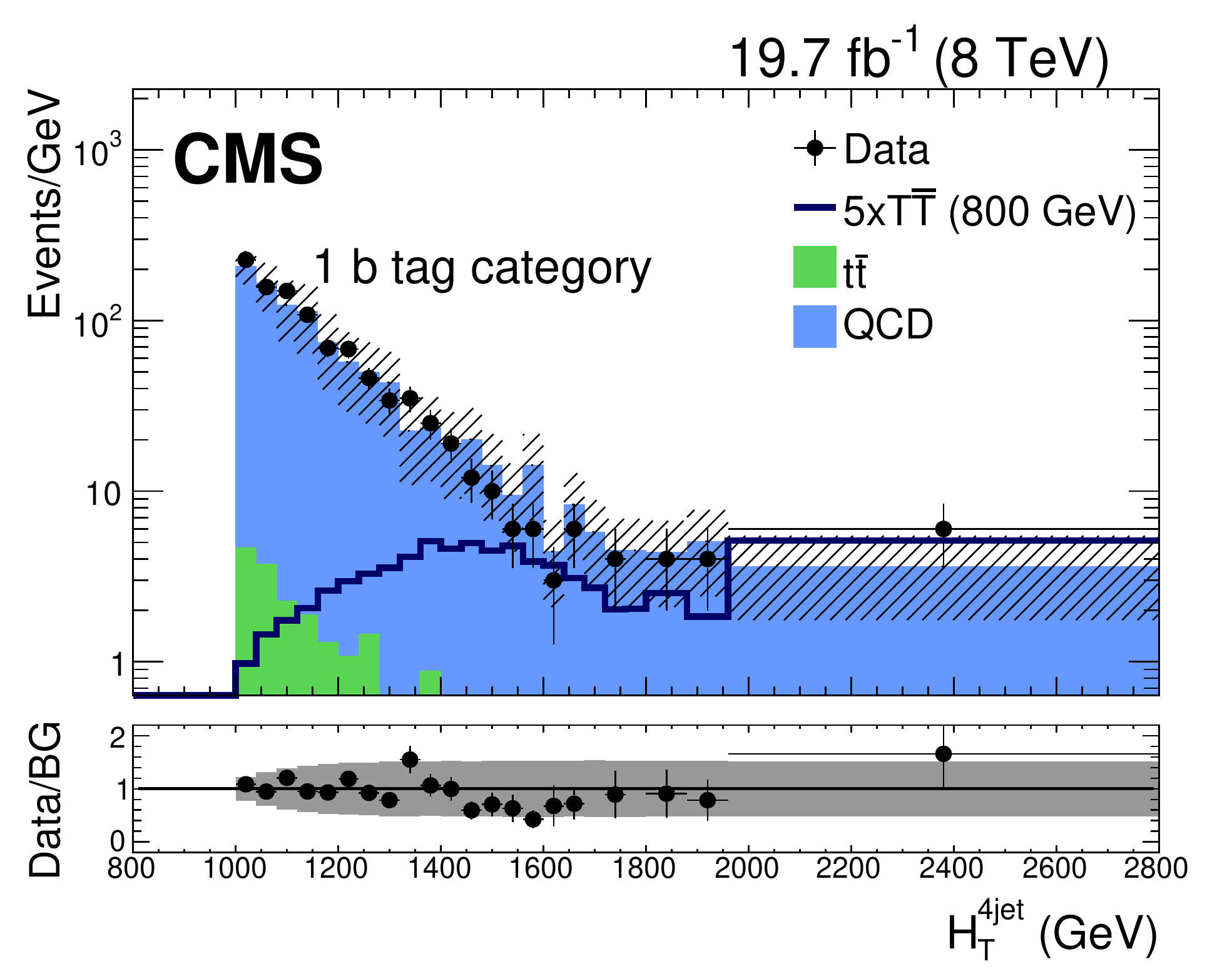}
\includegraphics[width=0.495\textwidth]{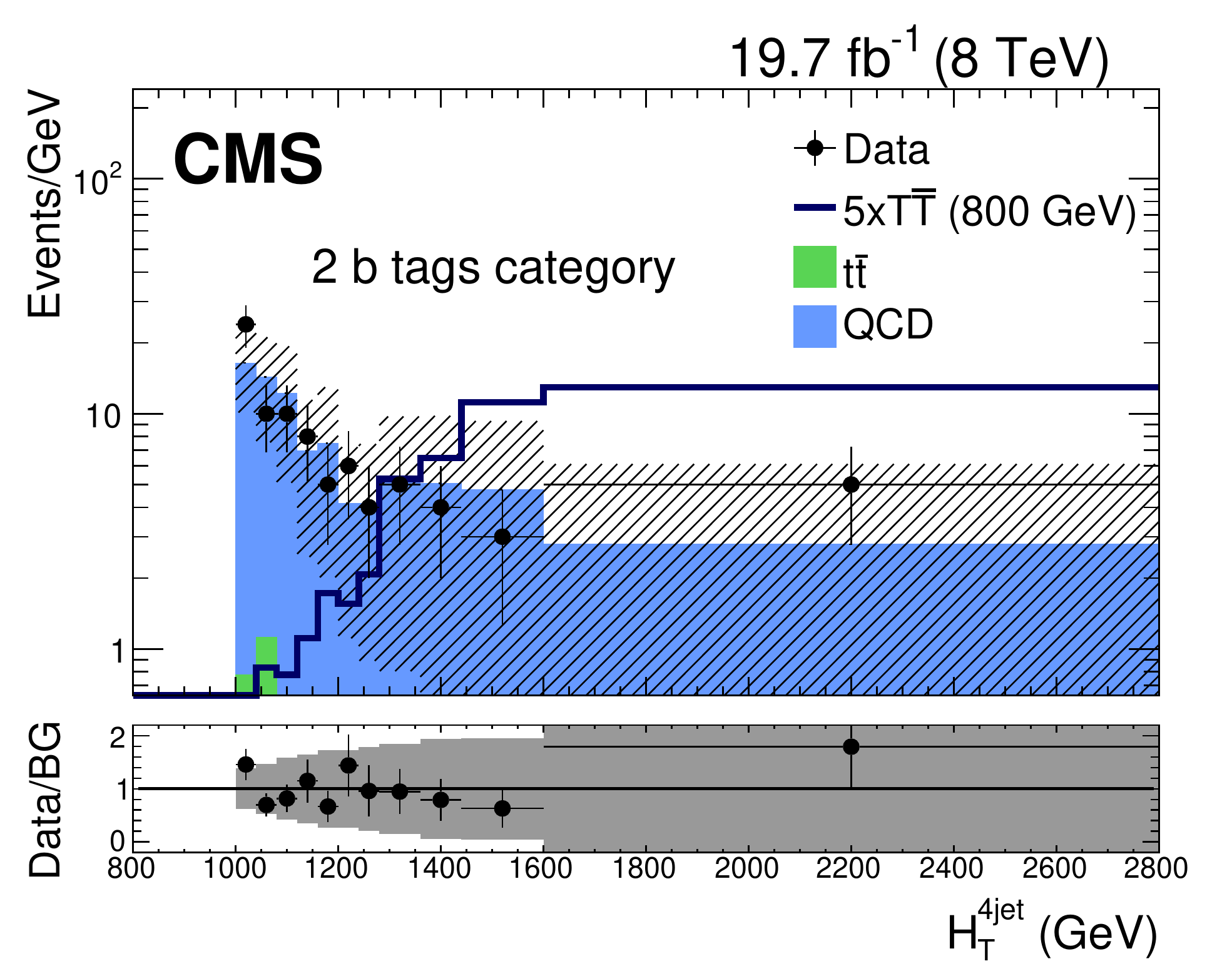}
\caption{The $\HT^\text{4jet}$ distributions for single \PQb tag events (left) and for events with at least two \PQb tags (right) for the all-hadronic search for $ \cPQT\to \bW$, including the QCD multijet background estimate obtained from data and the \cPQT quark signal with a mass of 800\GeV. The hatched region and the shaded area in the lower panel represent both the statistical and the systematic uncertainties in the total background. The lower panel represents the ratio between data and the sum of the backgrounds (BG). The horizontal error bars represent the bin width.}
\label{fig:ST_allhad}
\end{figure*}

Figure \ref{fig:ST_allhad} shows the post-fit $\HT^text{4 jet}$ distributions obtained with the above method.  Data are found to be in agreement with the expected background contributions. The numbers of expected background events and events observed in data after full selection are shown in Table~\ref{tab:yields}. The numbers of expected signal events and selection efficiencies assuming $\mathcal{B} (\cPQT \to \bW) = 100\%$ are summarized in Table~\ref{tab:signal12013}.

\begin{table}[h]
    \centering
  \topcaption{Summary of expected and observed background yields for the two channels of the $ \cPQT\to \bW$ search in the all-hadronic final state.}
    \label{tab:yields}
    \begin{scotch}{lcc}
        & 1 \PQb tag channel & $\ge$2 \PQb tags channel \\
     \hline
    \ttbar & $20.3\pm1.3$  &  $3.45\pm0.55$ \\
    QCD multijet    & $979\pm29$  & $80.2\pm6.4$  \\
    \hline
    Total background  & $999 \pm 31$   &  $84 \pm 7$   \\
    Data   & $998$  &  $84$ \\
    \end{scotch}
\end{table}

\begin{table*}[htb]
\topcaption{Selection efficiencies and numbers of expected signal events, for the two channels of the $ \cPQT\to \bW$ search in the hadronic final state. Different \cPQT quark mass hypotheses are considered and a 100\% branching fraction to \bW is assumed.}
\label{tab:signal12013}
\centering
\begin{scotch}{ccccc}
\cPQT quark mass    &\multicolumn{2}{c}{1 \PQb tag channel} &\multicolumn{2}{c}{$\ge$2 \PQb tags channel}\\
$(\GeVns)$ & Efficiency & Events & Efficiency & Events \\
\hline
500  & $1.01\%$ & $103.4$& $0.86\%$ & $84.7$ \\
600  & $2.24\%$ & $66.0$ & $1.81\%$ & $52.5$ \\
700  & $3.15\%$ & $31.24$ & $2.35\%$ & $22.80$ \\
800  & $4.07\%$ & $14.64$ & $2.51\%$ & $8.79$ \\
900  & $4.68\%$ & $6.57$ & $2.44\%$ & $3.33$ \\
1000 & $4.95\%$ & $2.81$ &$2.35\%$ & $1.29$ \\
\end{scotch}

\end{table*}

\subsection{Search for \texorpdfstring{$ \cPQT\to \tH$}{T to tH} with \texorpdfstring{$\PH \to \cPgg \cPgg$}{H to gamma gamma} \label{B2G14003}}
The analysis described in this section is optimized for events with one \cPQT quark decaying to \tH, where the Higgs boson decays into a pair of photons. The main advantage of this channel is  the possibility to precisely measure the invariant mass
of the diphoton system ($m_{\cPgg\cPgg}$) so that a peak in the $m_{\cPgg\cPgg}$ distribution would be present for signal events. The disadvantage is the small Higgs branching fraction of the order of $2 \times 10^{-3}$~\cite{Heinemeyer:2013tqa}.  The analysis concept is the same as  for searches of the  SM Higgs boson
in the $\PH \rightarrow\cPgg\cPgg$ decay channel \cite{Khachatryan:2014ira}.

Events with two isolated photons are selected.
Additional leptons and jets coming from the decay
of top quarks or a second Higgs boson are required.
In order to maximize the sensitivity of the analysis, two search channels are defined, targeting different decay modes of the top quark:
\begin{itemize}
\item the leptonic channel searches for events with a pair of photons and at least one isolated high-$\pt$ muon or electron;
\item the hadronic channel searches for events with a pair of photons and no isolated muons or electrons.
\end{itemize}

The resonant contributions from the $\ttbar \PH$  background are determined from simulation.
The nonresonant contribution is composed of events with two prompt photons arising from QCD multijet production as well as for emission in top quark production
($\cPgg\cPgg$+jets, $\ttbar+\cPgg\cPgg$, $\PQt+\cPgg\cPgg$). The \ttbar events are more likely to have a jet misreconstructed as a photon, because of the large numbers of jets in the final state. The simulation  of such sources of
instrumental background is not completely reliable. The background model is therefore derived from data.

The control sample used to estimate the nonresonant background consists of events where at least one photon passes
loose identification requirements but does not pass the final event selection. This sample is enriched with
 events containing one misidentified photon. A reweighting is applied, in order to match the \pt and $\eta$ spectra of the photons in this control sample to those obtained after the signal selection. This is done independently for each photon.

The event selection is based upon six quantities that have the largest
discriminating powers between signal and backgrounds and that have small correlations.
They include the transverse momenta of the larger \pt photon ($\cPgg_{1}$) and smaller \pt
photon ($\cPgg_{2}$).  The selection criteria are optimized to produce the most stringent limits on the signal
cross section and are listed Table~\ref{tab:selection} for both leptonic and hadronic channels.

\begin{table}[htb]
\centering
\topcaption{Final selection criteria for hadronic and leptonic channels of the search for $ \cPQT\to \tH$ with $\PH \to \cPgg \cPgg$.
\label{tab:selection}}
\begin{scotch}{ccc}
Variable & Leptonic channel& Hadronic channel \\
\hline
$\pt(\cPgg_{1})$ & ${>}\frac{1}{2}\mgg$& ${>}\frac{3}{4}\mgg$ \\
$\pt(\cPgg_{2})$ & 25\GeV & 35\GeV \\
Number of jets &  $\ge$2 & $\ge$2 \\
\sT & $\ge$770\GeV & $\ge$1000\GeV \\
Leptons & $\ge$1 & 0 \\
\PQb tags & \NA & $\ge$1 \\
\end{scotch}
\end{table}

\begin{figure*}[htb]
\centering
\includegraphics[width=0.4\textwidth]{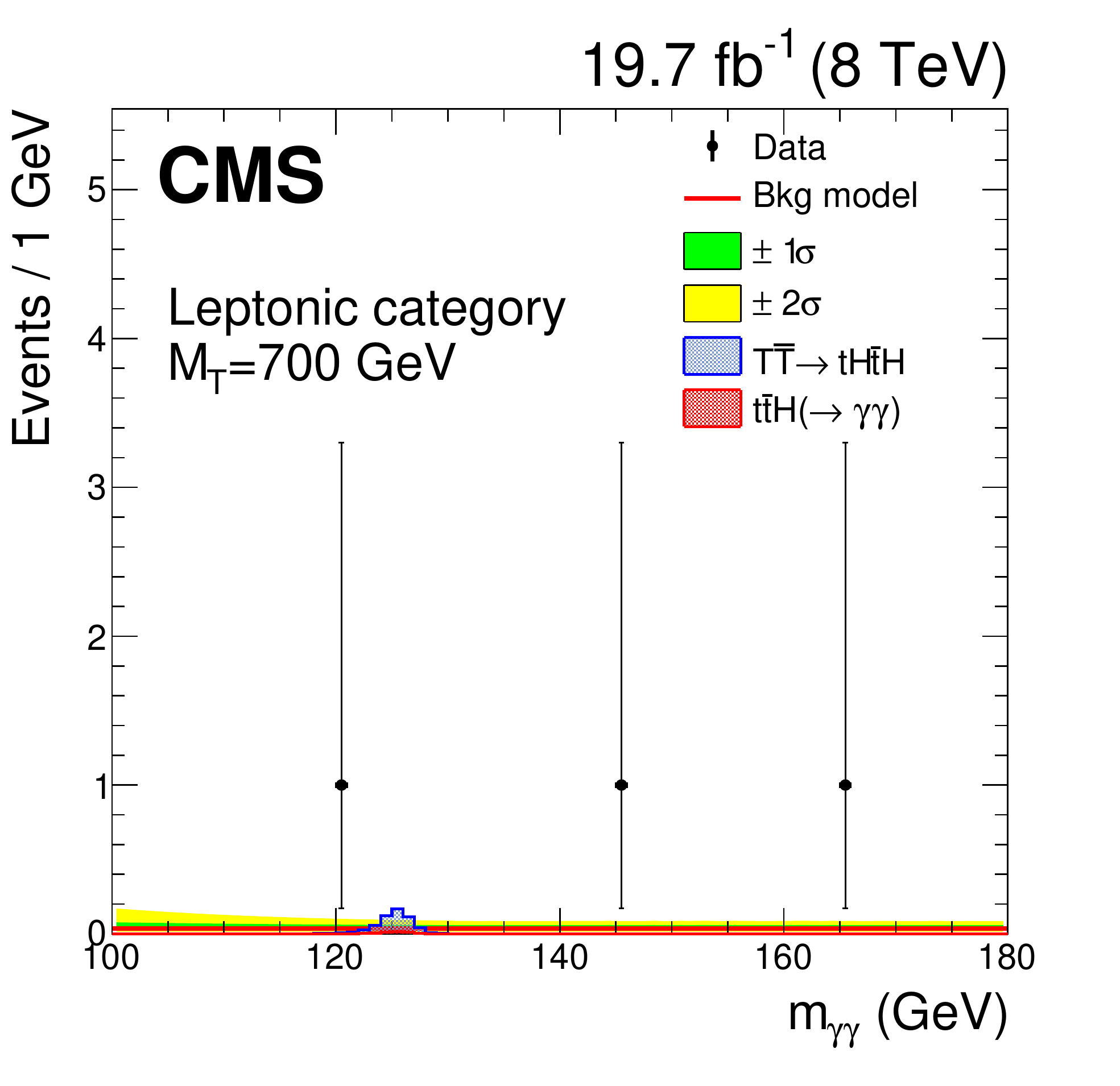}
\includegraphics[width=0.4\textwidth]{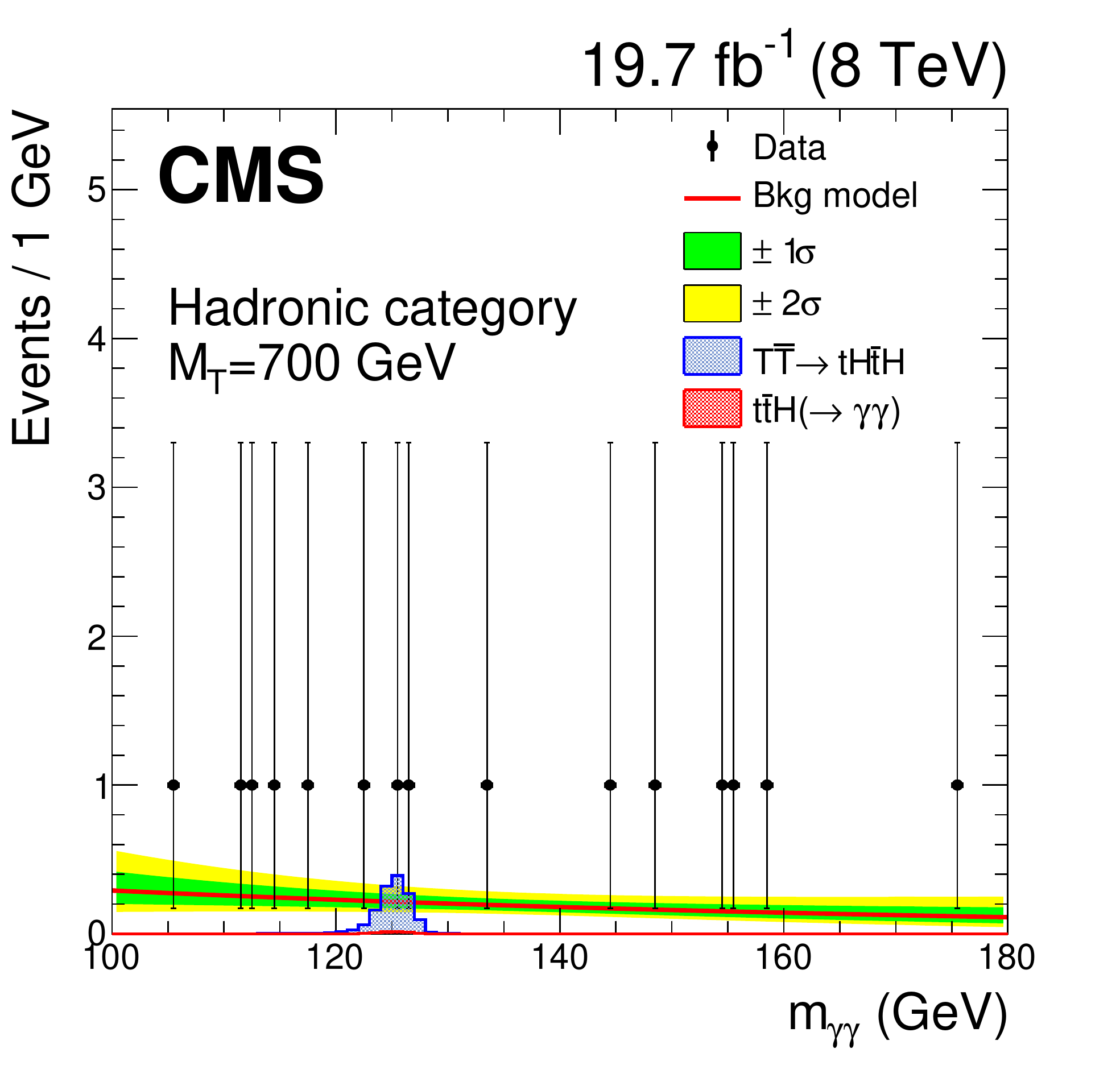}
\caption{Diphoton invariant mass distribution for the leptonic (left) and hadronic (right) channels of the search for $ \cPQT\to \tH$ with H$\to \cPgg \cPgg$. The signal is normalized to the
predicted theoretical cross section corresponding to $m_{\mathrm{T}}=700\GeV$.  The backgrounds predicted  by the fit are shown as a solid line while the corresponding uncertainties are shown as bands around the line, where the inner band indicates the  $1\sigma$  and the outer band indicates the $2\sigma$ uncertainties. Bins with zero entries are not shown.  \label{fig:Hggfit}}
\end{figure*}

The nonresonant background contributions are obtained from unbinned maximum likelihood fits to the diphoton mass distribution
over the range $ 100 <m_{\cPgg\cPgg}< 180\GeV$, under the hypothesis of no signal. An exponential function is chosen
for these fits. Studies of pseudo-experiments showed that the use of an exponential function does not introduce a bias
in the estimation of the numbers of background events in both categories.
In Fig. \ref{fig:Hggfit}, the observed diphoton mass distribution in each event category is shown,
 together with the expected signal and the expected resonant background contribution.
The error bands show the uncertainty in the background shapes associated with the statistical uncertainties of the fits. The numbers of expected background events and events observed in data after final selection are shown in Table~\ref{tab:yieldsgamma}. The numbers of expected signal events and selection efficiencies assuming $\mathcal{B} (\cPQT \to \tH) = 100\%$ are summarized in Table~\ref{tab:signal14003}.

 \begin{table}[htb]
   \topcaption{Expected yields for $\ttbar \PH$ and nonresonant background (from the fit to data) and the numbers of observed events in data after full
event selection for the two channels of the $ \cPQT\to \tH$ search in the final state with photons. All the yields are computed in a window of 1 full width at half maximum \ie, $125\pm1.5\GeV$.
    \label{tab:yieldsgamma}}
    \centering
      \begin{scotch}{lcc}
         & Leptonic channel& Hadronic channel \\
\hline\\[-0.3cm]
        $\ttbar \PH$ & 0.039$^{+0.005}_{-0.006}$ & 0.042$^{+0.005}_{-0.006}$ \\[0.2cm]
      Nonresonant  background & $0.11^{+0.07}_{-0.03}$ & $0.65^{+0.16}_{-0.13}$ \\[0.2cm]
\hline      \\[-0.3cm]
Total background & $0.15^{+0.07}_{-0.03}$ & $0.69^{+0.16}_{-0.13}$ \\
        Data & 0 & 2 \\
      \end{scotch}
  \end{table}

\begin{table*}[htb]
\topcaption{Selection efficiencies and numbers of expected signal events, for the two channels of the $ \cPQT\to \tH$ search in the final state with photons. Different \cPQT quark mass hypotheses are considered and a 100\% branching fraction to \tH is assumed.}
\label{tab:signal14003}
\centering
\begin{scotch}{ccccc}
\cPQT quark mass    &\multicolumn{2}{c}{Leptonic channel} &\multicolumn{2}{c}{Hadronic channel}\\
$(\GeVns)$ & Efficiency & Events & Efficiency & Events \\
\hline
500  & $6.7\%$ & 6.0 & $9.3\%$ & 8.3 \\
600  & $9.6\%$ & 8.7 & $18.1\%$ & 16.4 \\
700  & $11.0\%$ & 9.8 & $26.0\%$ & 23.8 \\
800  & $12.0\%$ & 10.9 & $30.0\%$ & 27.3 \\
900  & $11.4\%$ & 10.4 & $32.0\%$ & 29.3 \\
\end{scotch}

\end{table*}

The data in the signal window are compatible with background expectations from SM processes.

\section{Combination strategy \label{sec:combo}}

The event samples selected by the five analyses are almost entirely distinct and therefore, signal limits extracted from those analyses are statistically independent.  They can be combined to yield a result that is more stringent than any of the inputs.  Because the backgrounds are largely common to all analyses, the background estimates are largely correlated but well determined by the multiple independent samples.  In particular, most analyses have top quark pair production as a background process. This background normalization is correlated among the analyses in the combination, providing for the
combination a better background estimation than in the individual
analyses. Similar arguments hold for the correlated systematic uncertainties, which are discussed in more detail in Section  \ref{sec:systematics}.

The  inclusive analysis with single and multiple leptons described in
Section \ref{B2G12015} is able to set limits for all \cPQT quark decay modes. Dedicated optimizations to enhance the sensitivity for $\cPQT \to
\bW$ decays are described in Section \ref{B2G12017}. These
optimizations use single-lepton events. To avoid double counting of events we
 replace the single-lepton part of the inclusive approach (Section
 \ref{B2G12015}) with the single-lepton analysis described in Section \ref{B2G12017}. This is done for scenarios with $\mathcal{B}(\cPQT \to \bW)$ values of at least 80\%. For lower $\mathcal{B}(\cPQT \to \bW)$ values this
approach is inferior and  we use the inclusive results from
Section \ref{B2G12015} only. At every point the approach used  is that which gives the best expected
limit.
The other three analyses described in Sections \ref{B2G14002} to \ref{B2G14003} do not
have any overlap so  they are always combined
with the cases above.

For the statistical combination a Bayesian method \cite{bayesian} has been adopted in
which the systematic uncertainties are taken into account as nuisance
parameters with their corresponding priors as discussed in Section
\ref{sec:systematics}.  Upper limits on the \cPQT quark production cross
section  are obtained with the Theta
framework~\cite{theta_web}. Systematic uncertainties are taken into account as  global normalization
 uncertainties and as shape uncertainties where applicable. More
 details about the treatment of systematic uncertainties are given in
 the next section.

\subsection{Systematic uncertainties \label{sec:systematics}}

Some of the individual analyses are sensitive to the same systematic
uncertainties, for example the uncertainty in the integrated luminosity, the jet
energy scale and the \PQb tagging efficiency. Such uncertainties are
treated as fully correlated, as is done technically by correlating the
corresponding nuisance
parameters in the limit setting procedure. This treatment allows improved constraints to be obtained on these parameters than is possible in the
standard analyses.

The systematic uncertainties fall into two types: those which affect
the normalization of the signal and background samples, and those
which also affect the shapes of distributions.  The uncertainty in the
\ttbar cross section is 13\%. It is obtained  from the \ttbar cross section
  measurement~\cite{Khachatryan:2015oqa} for large invariant mass
  values of the \ttbar system.  The uncertainty in the integrated luminosity is 2.6\%~\cite{CMS-PAS-LUM-13-001}.

Shape uncertainties
include  the jet energy scale, the jet energy
resolution and the \PQb tagging
efficiency uncertainties. We also consider the uncertainties in the
efficiencies of the \PQt tagging, \PW tagging, and
\PH tagging  algorithms \cite{CMS-PAS-JME-13-007,JME13006,CMS-PAS-BTV-13-001}.
 The uncertainty due to the energy deposits not associated with jets (unclustered energy) has an impact on the missing \pt. This effect is taken into account in the single-lepton channel. The size of this uncertainty typically varies from a few percent up to 10\%.

The systematic uncertainty in the pileup jet identification is taken into
account in the analysis with $\PH \to \cPgg \cPgg$. It is derived through the use of the data/simulation scale
factors (SF), which are binned in jet $\eta$ and \pt~\cite{Khachatryan:2014ira}.

For the photon identification efficiency,   the
uncertainty in the SF is taken into account. The SF corrects the
efficiency in simulation to the efficiency as measured in data  using a ``tag-and-probe''
technique~\cite{CMS:tagandprobe} applied to $\cPZ \to \Pe^{+}\Pe^{-}$ events. The uncertainty applied to this SF amounts to 3\% in the barrel region of the calorimeter and 4\% in the endcaps.

Lepton trigger efficiencies, lepton identification efficiencies, and corresponding correction  factors for simulated events are obtained from data using decays of \cPZ bosons to dileptons. These uncertainties are $\le 3\%$.

For simulated $\ttbar$ and  $\PQt \PQt \PH$ events, uncertainties due to
renormalization and factorization scales ($\mu_\mathrm{R}$ and $\mu_\mathrm{F}$) are taken
 into account by varying both scales simultaneously up and down by a
 factor of two. Uncertainties
 arising from the choice of PDFs are taken into account. Simulated background events are weighted
  according to the uncertainties parameterized by the CTEQ6
  eigenvectors \cite{1126-6708-2002-07-012}. The shifts produced by
  the individual eigenvectors are added   in quadrature in each bin of the relevant distributions.

A systematic uncertainty of 50\% is assigned to the diboson backgrounds, single top quark production and
the \PW and \cPZ boson background. This accounts for the effects of the $\mu_\mathrm{R}$ and $\mu_\mathrm{F}$ variations in simulation
and the uncertainties in the determination of the $\PW$+jets SF from data.

Modified ``template''  distributions of those quantities that are
affected by the respective uncertainties are obtained by varying the
respective quantity by its uncertainty, namely by $\pm$1 standard deviation.
In the limit setting procedure  a likelihood fit is performed in
which the nominal distribution and the modified  templates are
interpolated.  The
corresponding uncertainty is represented as a nuisance parameter, which receives
its prior constraints from the template distributions. In the fit, the templates are allowed to be extrapolated beyond
$\pm$1 standard deviation, but this happens rarely. The resulting fit
values are always within $\pm$1.5 standard deviations of their prior
values.

The list of nuisance parameters of all analysis channels is shown in Table
\ref{tab:nuisance_table}. This table also indicates which parameters are correlated and which
uncorrelated.

\begin{table*}[htbp]
\topcaption{Correlated and uncorrelated systematic uncertainties. The
 \checkmark \ symbol  indicates that the uncertainty has been taken
  into account in the analysis, but it is not correlated with any of
  the other analyses. The \boxCheck \ symbol  indicates that the
  uncertainty has been taken into account and that it is correlated
  with the other analysis that have a \boxCheck \  sign as well. A
  missing symbol indicates that this uncertainty is not relevant for
  this analysis channel.}
\centering
\begin{scotch}{lcccccc}
                                                             &Single       & Inclusive       & Multiple       &     All-had.       & All-had.       & $\PH \to \cPgg\cPgg$ \\
Uncertainty                                           &  leptons   &  leptons        & leptons         &   $ \cPQT\to \bW$      &  $ \cPQT\to \tH $      &              \\
\hline
Int. luminosity                                           & \boxCheck& \boxCheck   & \boxCheck   & \boxCheck       & \boxCheck    &    \boxCheck    \\
Trigger                                                 & \checkmark&   \checkmark & \checkmark&                   & \checkmark  & \checkmark     \\
Lepton ID                                             &\boxCheck & \boxCheck  & \boxCheck &                         &                      & \boxCheck    \\
Photon ID                                             &                  &                     &                     &                          &                      & \checkmark    \\
Photon energy                                      &                  &                     &                     &                          &                      & \checkmark    \\
Pileup jet ID                                         &                  &                     &                     &                          &                      & \checkmark    \\
Jet energy scale                                   & \boxCheck& \boxCheck    & \boxCheck   & \boxCheck      &  \boxCheck                      &  \boxCheck\\
Jet energy resolution                           & \boxCheck& \boxCheck    &\boxCheck    & \boxCheck      &   \boxCheck                     & \boxCheck \\
Unclustered energy                              &\checkmark&                       &                     &                         &                      & \\
\PQb tag SF                                              &\boxCheck & \boxCheck    & \boxCheck   &  \boxCheck      & \boxCheck    &   \\
\PQb tag mistag SF                                   &\boxCheck&                       &                     &                         & \boxCheck    &   \\
\PQt tagging SF                                    &                 &                       &                     &                         & \checkmark  &    \\
$\ttbar$ $\mu$R and $\mu$F scale           & \boxCheck& \boxCheck   & \boxCheck  & \boxCheck       & \boxCheck    &    \\
$\ttbar$ cross section       & \boxCheck& \boxCheck    & \boxCheck   & \boxCheck       & \boxCheck    &   \\
$\ttbar$  PDF                    & \boxCheck&  &   & \boxCheck     & \boxCheck    &   \\
QCD background                                &                  &                       &                     &     \checkmark & \checkmark    & \\
Other backgrounds                             &\checkmark & \boxCheck   & \boxCheck    & \checkmark      &                        & \checkmark \\
\end{scotch}
\label{tab:nuisance_table}

\end{table*}

\section{Results \label{sec:results}}
\begin{figure*}[htbp]
\centering
\includegraphics[width=0.49\linewidth]{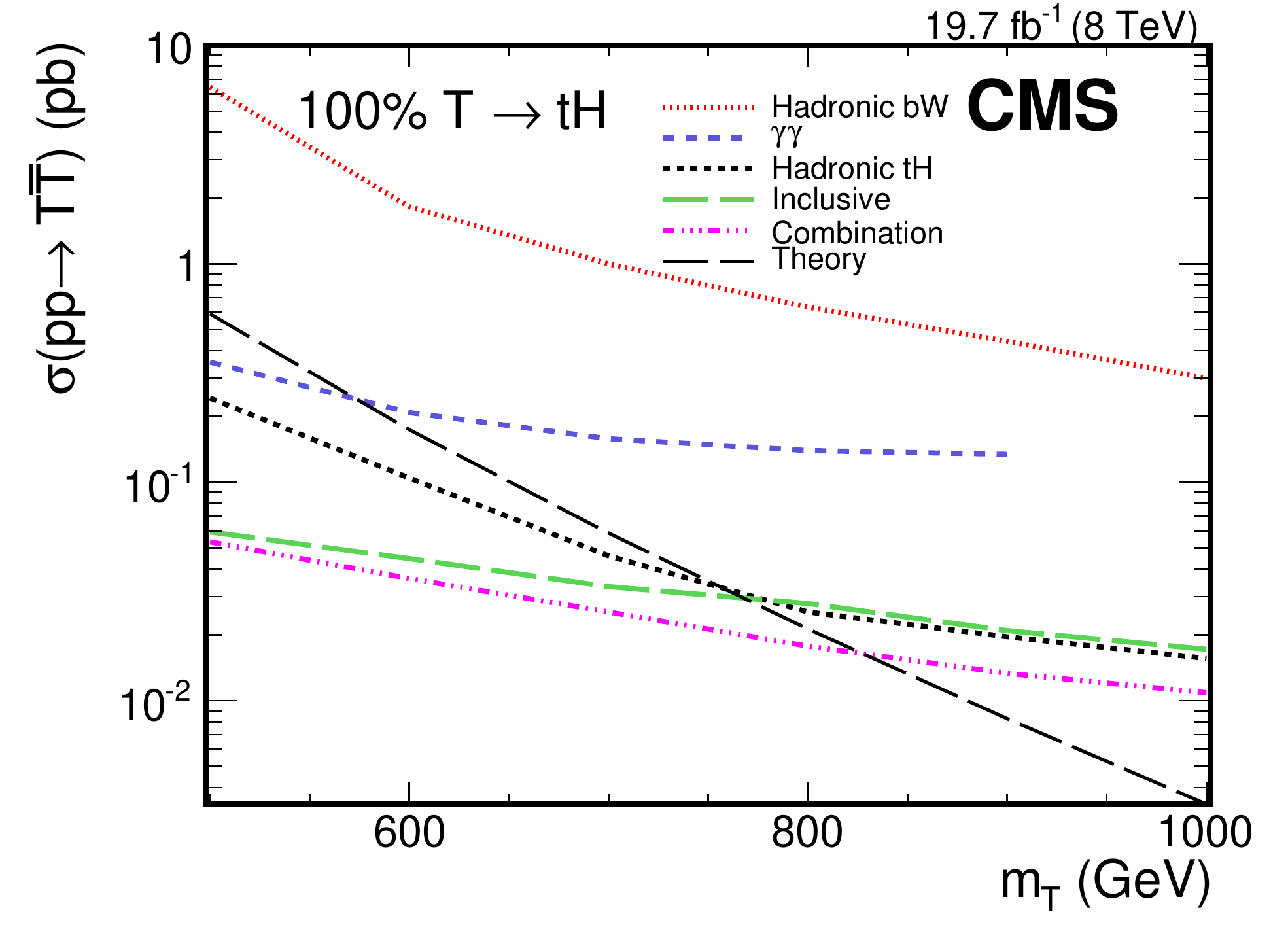}
\includegraphics[width=0.49\linewidth]{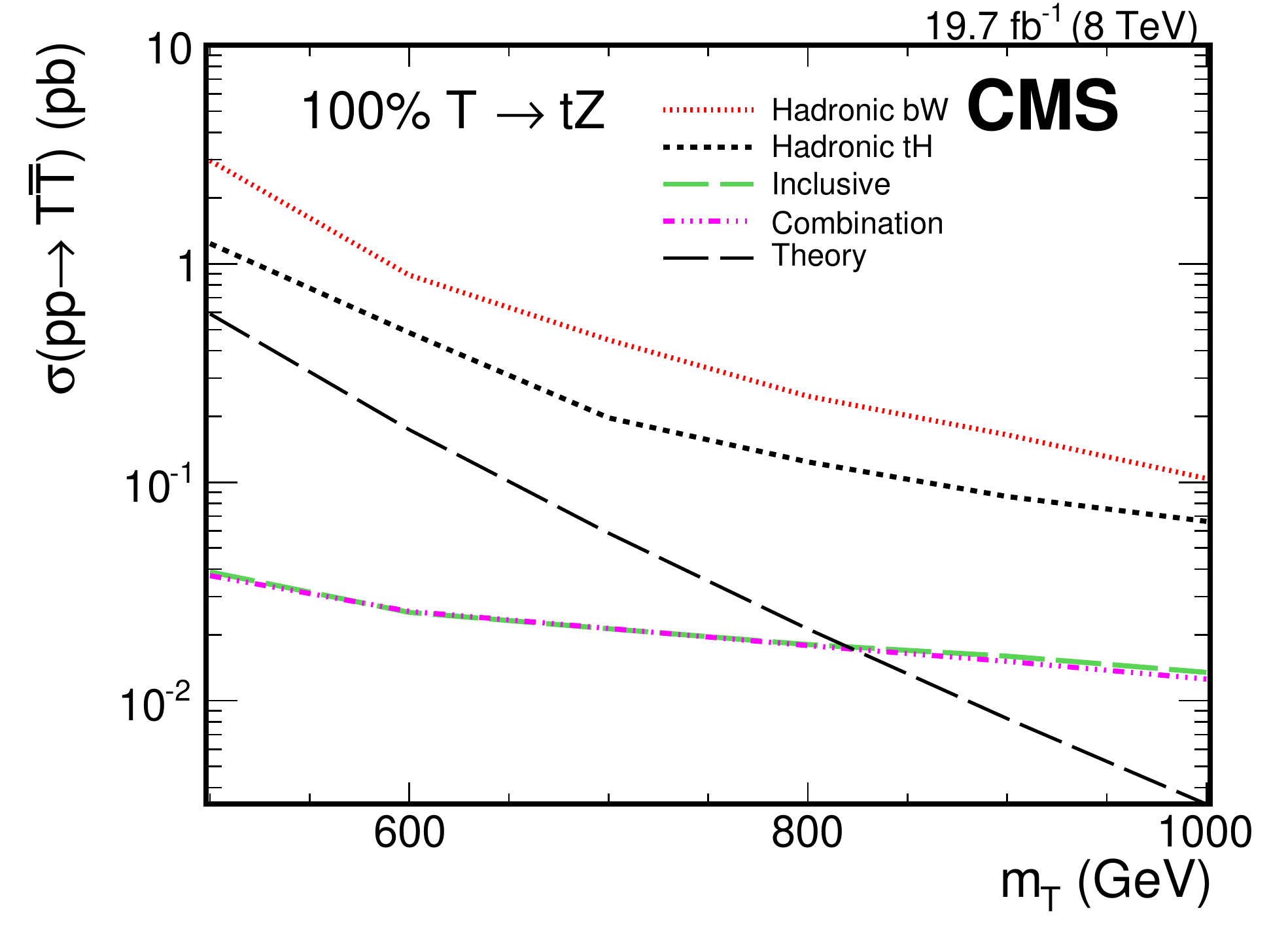}
\includegraphics[width=0.49\linewidth]{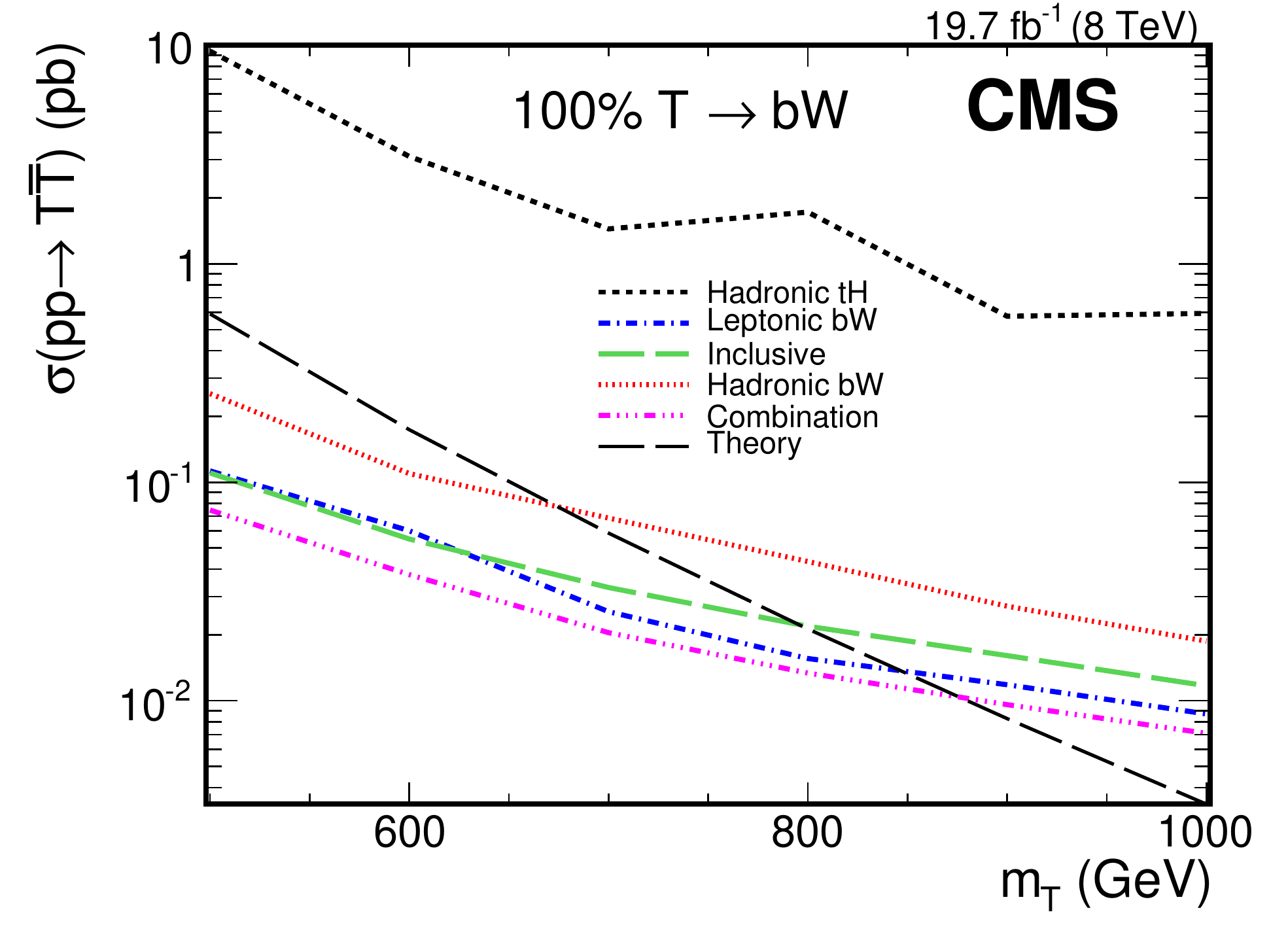}
\caption{Expected limits at 95\% CL of the individual analyses in comparison to
  the combination for exclusive decays of the \cPQT quark to
  \tH, \tZ, and \bW.}
\label{fig:compareExpected}

\end{figure*}

No significant deviation from the SM prediction is observed. The
expected  limits
of the individual analysis channels at a 95\% confidence level (CL) are displayed in
Fig.~\ref{fig:compareExpected} for  exclusive decays of the \cPQT quark to \tH, \tZ, and \bW. This figure also shows the result of the
combination,
where only the non-overlapping part of the individual analyses are
combined, as discussed in Section~\ref{sec:combo}.

\begin{figure*}[htbp]
\centering
\includegraphics[width=0.49\linewidth]{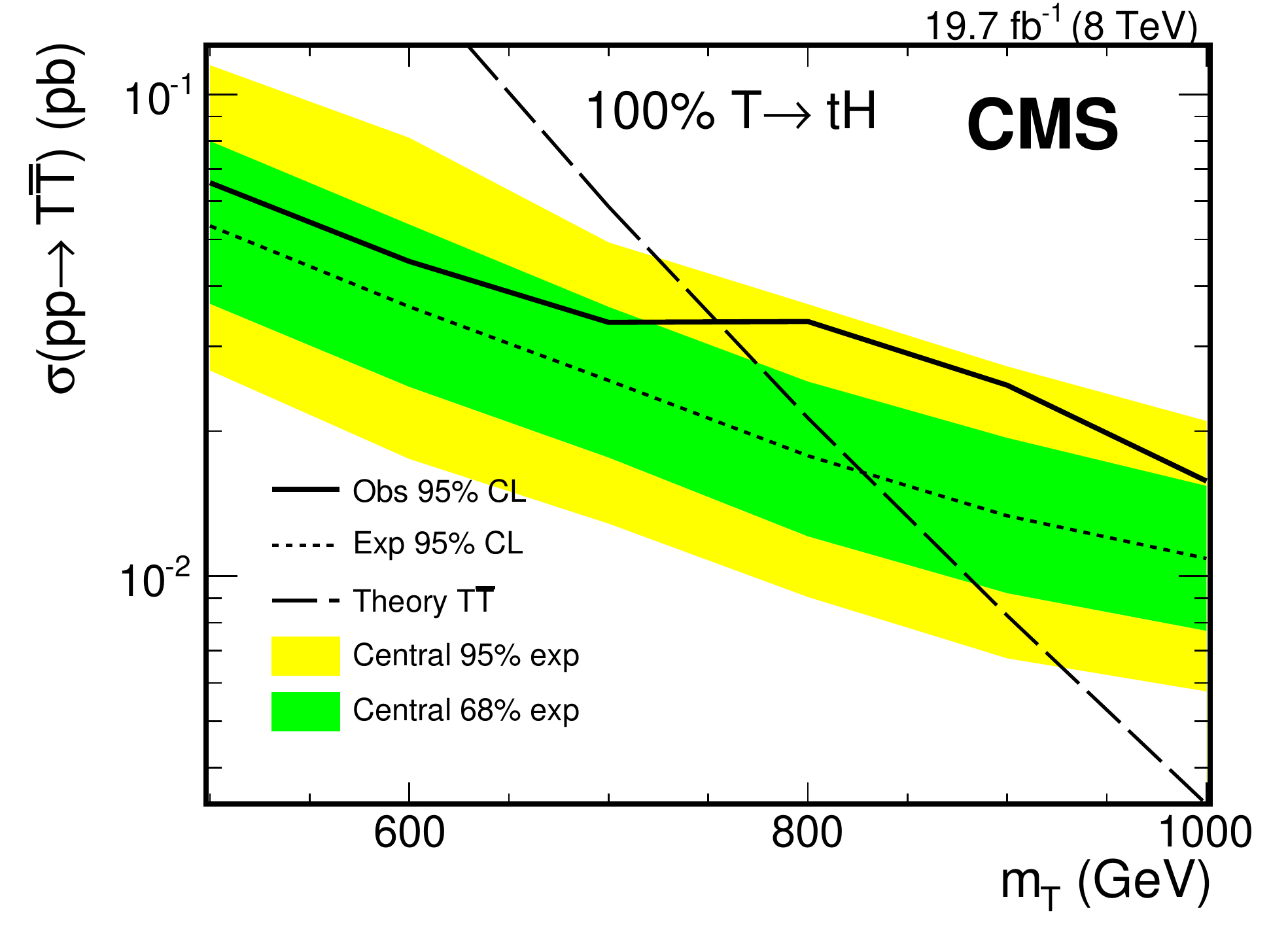}
\includegraphics[width=0.49\linewidth]{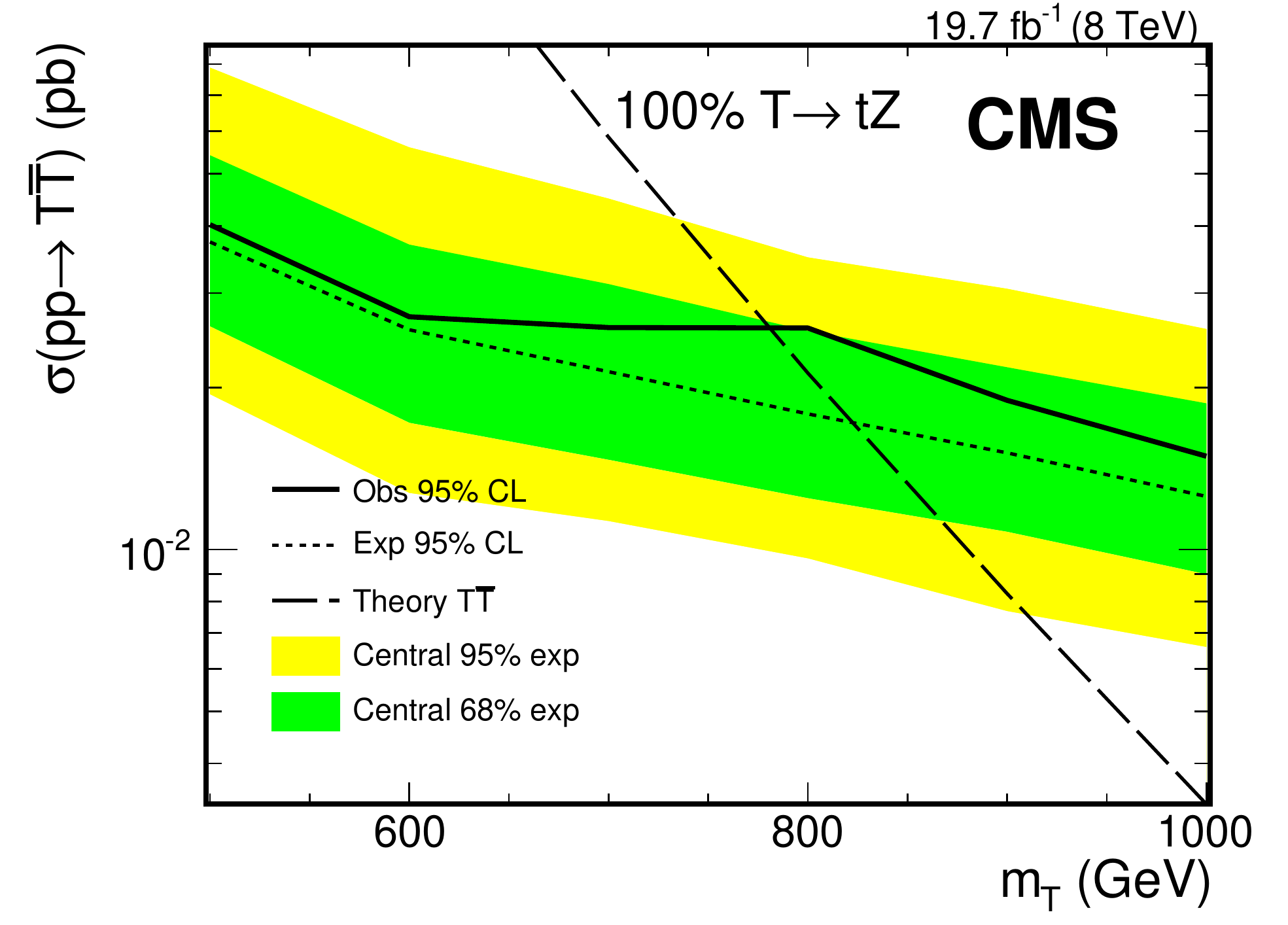}
\includegraphics[width=0.49\linewidth]{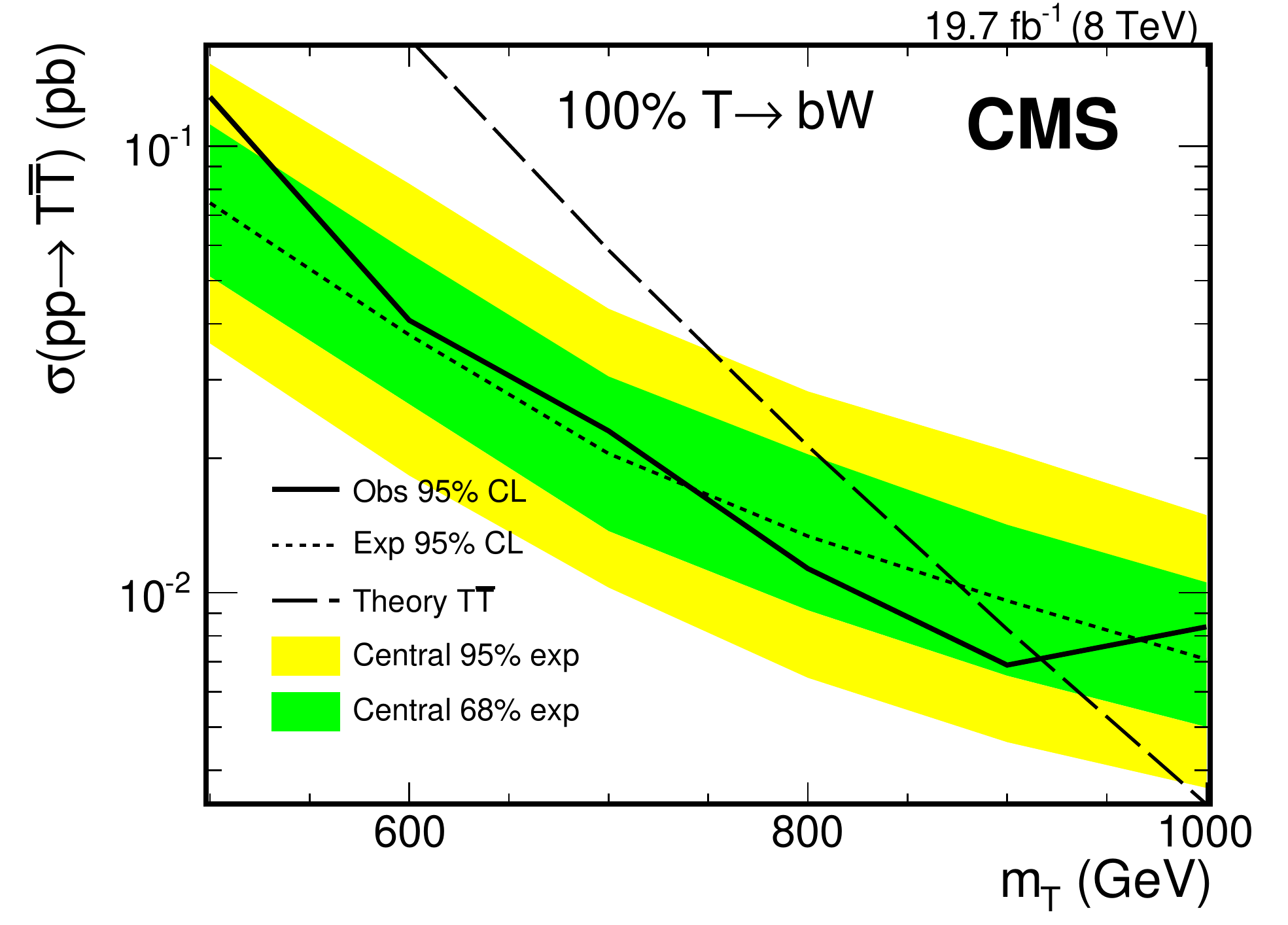}
\caption{Observed and expected Bayesian  upper limits  at 95\% CL on the \cPQT quark production
  cross section for  exclusive \cPQT quark decays to \tH, \tZ, and \bW.  The green (inner)
          and yellow (outer) bands show the   $1 \sigma$ ($2 \sigma$)
          uncertainty ranges in the expected limits, respectively. The
          dashed line shows the prediction of the theory. }
\label{fig:combinedExpectedObserved}

\end{figure*}

The  observed limits  and  the expected one and two standard deviation uncertainties are displayed in Fig. \ref{fig:combinedExpectedObserved} for exclusive \cPQT quark decays.

\begin{figure*}[htb]
\centering
\includegraphics[width=0.49\linewidth]{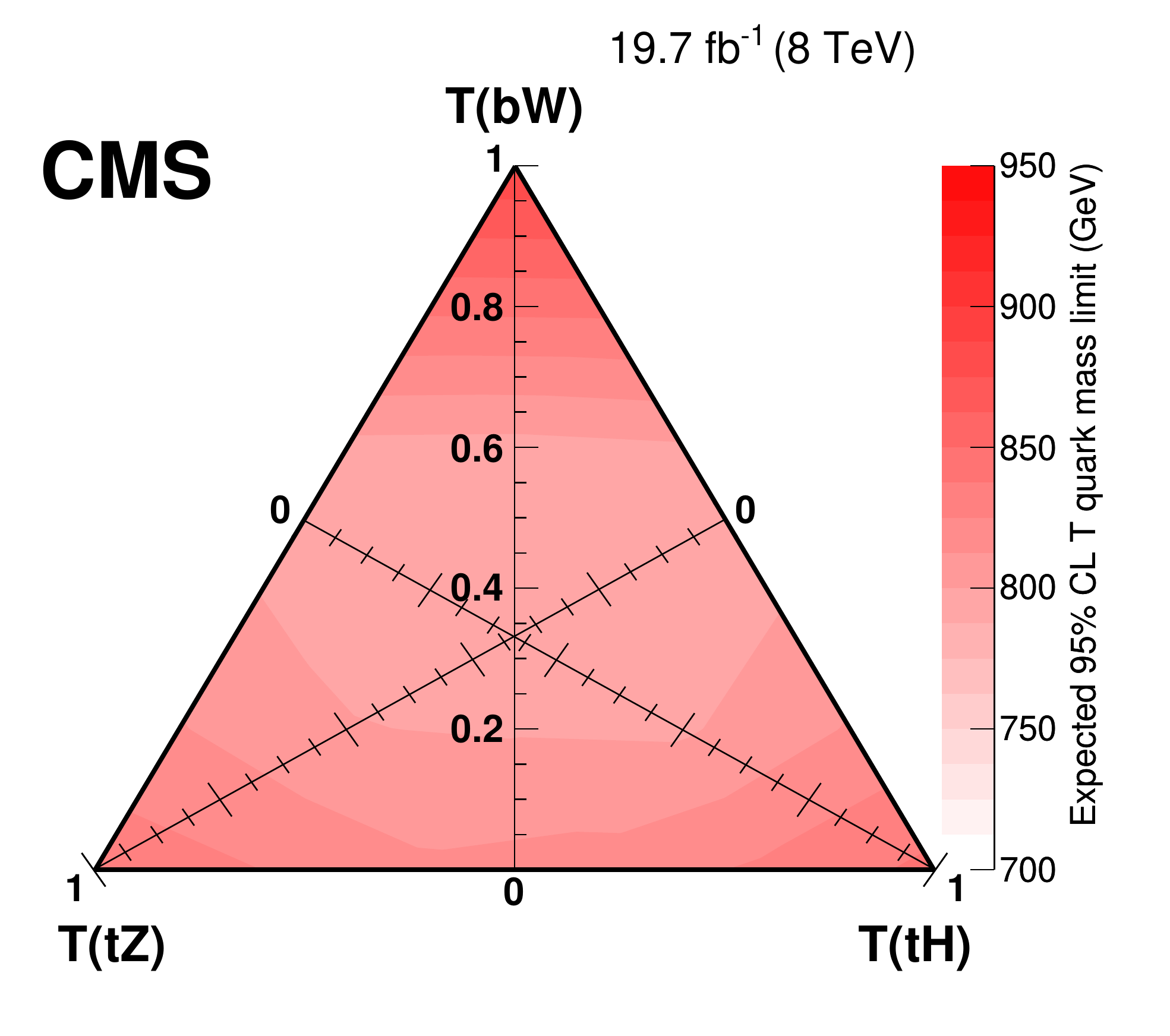}
\includegraphics[width=0.49\linewidth]{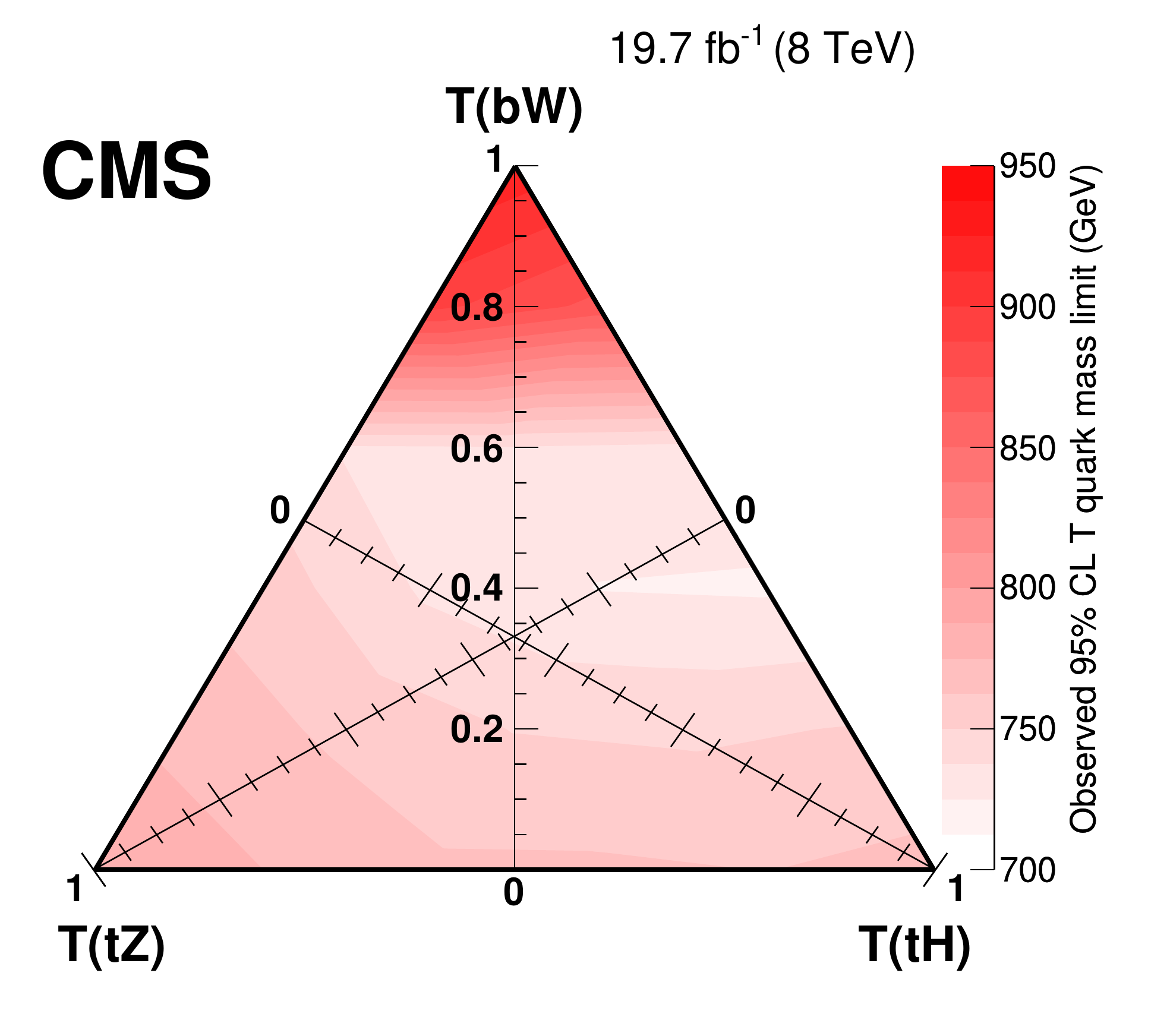}
\caption{Expected (left) and observed (right) 95\% CL limits of the combined
  analysis, visualized in a triangle representing the  branching
  fractions of the \cPQT quark decay.}
\label{fig:TriangleCombined}

\end{figure*}

\begin{figure*}[htb]
\centering
\includegraphics[width=0.49\linewidth]{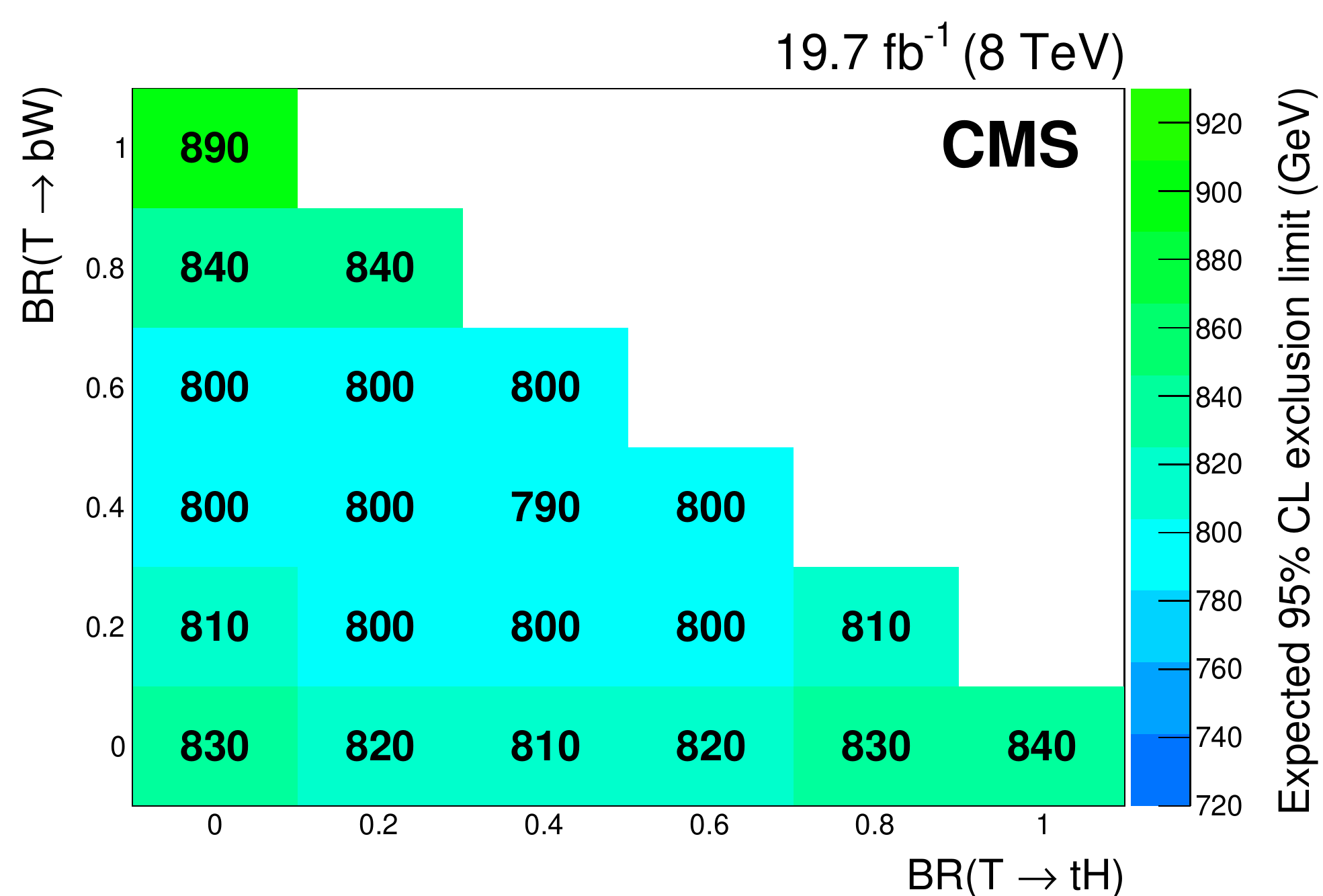}
\includegraphics[width=0.49\linewidth]{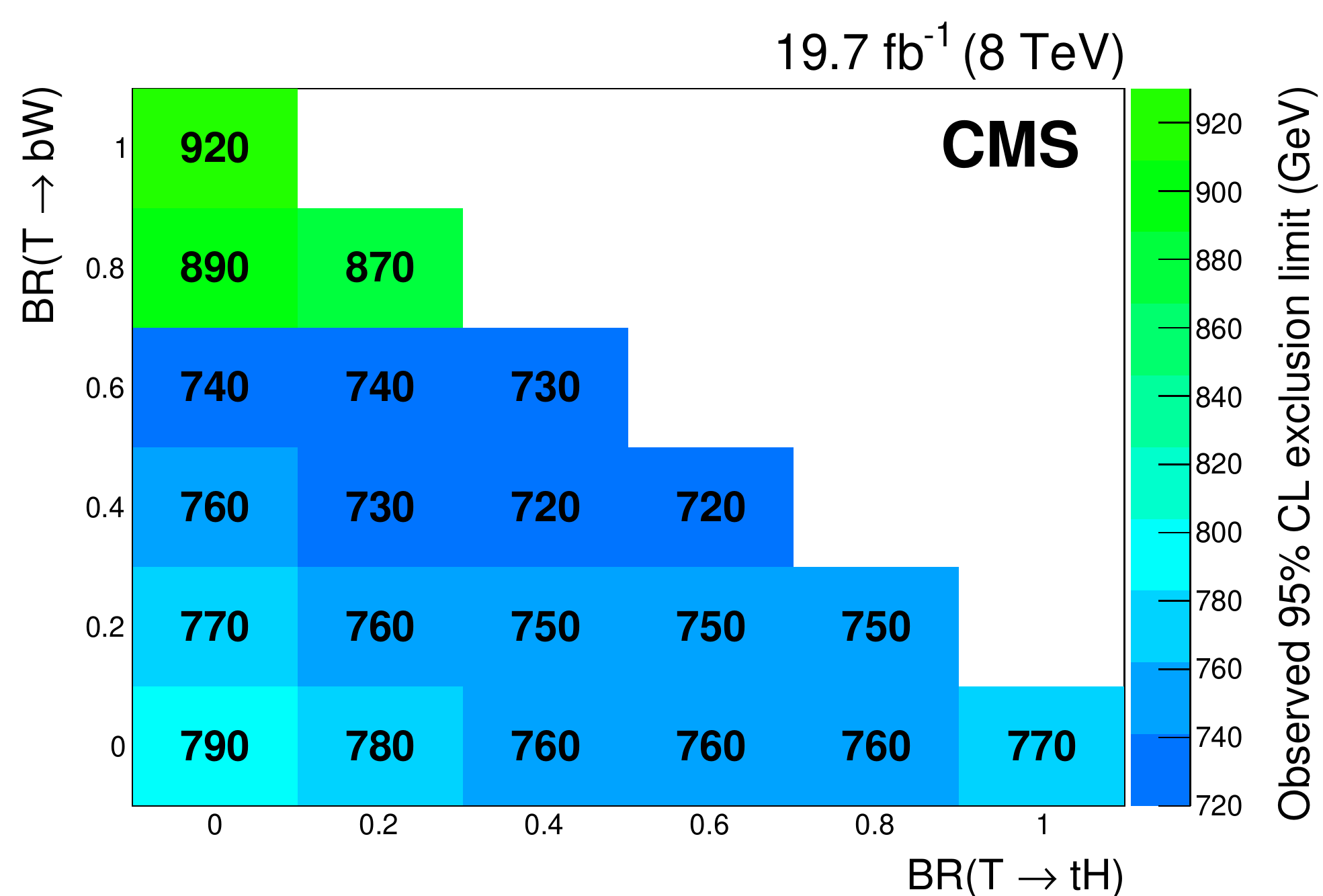}
\caption{Expected (left) and observed (right) 95\% CL limits of the combined
  analysis, for combinations of branching
fractions to \tH, \tZ, and \bW. The branching fraction to \tZ is not
explicitly reported, since it is  given by $1-\mathcal{B}(\tH)-\mathcal{B}(\bW)$.}
\label{fig:TriangleCombined2}
\end{figure*}

The lower  limits on the mass of the \cPQT quark are obtained by determining the intersection
between expected (observed) limits with the theoretical
prediction, based on the cross section versus \cPQT quark mass distributions shown in Fig.~\ref{fig:combinedExpectedObserved}. The results are visualized graphically in the triangular plane of branching fractions  in Fig.~\ref{fig:TriangleCombined}.  The numerical upper
limits on the \cPQT quark production cross section  are given in Table~\ref{branchingsInScanTableMass} for a full range  of branching fractions and the numerical results of the limits on the mass of the \cPQT quark are given in Table \ref{tab:limit_table}. A different visualization of the mass limits is presented in Fig.~\ref{fig:TriangleCombined2}.

\begin{table*}[htbp]
 \centering \topcaption{Branching fractions  (first three columns) and
   the observed and expected upper  limits on the \cPQT quark production cross section  at 95\% CL for different values of
  the \cPQT quark mass. The expected limits are quoted with their corresponding
  uncertainties, for different branching fractions hypotheses. The cross section
  limits are given in units of pb.}
\label{branchingsInScanTableMass}
\begin{scotch}{ccccccccc}
$\mathcal{B}$ & $\mathcal{B}$ & $\mathcal{B}$ & \multicolumn{6}{c}{\cPQT quark mass (\GeVns)} \\
(\tH) & (\tZ) & (\bW) &500 & 600 & 700 & 800 & 900 & 1000\\
\hline
0.0&1.0&0.0&0.037$^{+0.017}_{-0.011}$&0.026$^{+0.011}_{-0.009}$&0.021$^{+0.010}_{-0.006}$&0.018$^{+0.008}_{-0.006}$&0.015$^{+0.007}_{-0.004}$&0.013$^{+0.006}_{-0.004}$\\
&&&0.040&0.027&0.026&0.026&0.019&0.015\\
0.2&0.8&0.0&0.043$^{+0.022}_{-0.014}$&0.029$^{+0.013}_{-0.009}$&0.023$^{+0.012}_{-0.007}$&0.019$^{+0.009}_{-0.006}$&0.016$^{+0.008}_{-0.005}$&0.013$^{+0.005}_{-0.004}$\\
&&&0.045&0.030&0.031&0.030&0.023&0.016\\
0.4&0.6&0.0&0.049$^{+0.022}_{-0.016}$&0.033$^{+0.015}_{-0.011}$&0.025$^{+0.010}_{-0.008}$&0.020$^{+0.010}_{-0.006}$&0.016$^{+0.007}_{-0.005}$&0.013$^{+0.006}_{-0.004}$\\
&&&0.052&0.033&0.032&0.036&0.025&0.018\\
0.6&0.4&0.0&0.053$^{+0.025}_{-0.018}$&0.035$^{+0.015}_{-0.011}$&0.026$^{+0.012}_{-0.008}$&0.020$^{+0.009}_{-0.006}$&0.016$^{+0.006}_{-0.005}$&0.013$^{+0.005}_{-0.004}$\\
&&&0.066&0.038&0.033&0.035&0.024&0.017\\
0.8&0.2&0.0&0.055$^{+0.027}_{-0.018}$&0.036$^{+0.017}_{-0.011}$&0.026$^{+0.011}_{-0.009}$&0.019$^{+0.009}_{-0.006}$&0.015$^{+0.006}_{-0.005}$&0.012$^{+0.005}_{-0.004}$\\
&&&0.058&0.039&0.035&0.036&0.025&0.016\\
1.0&0.0&0.0&0.053$^{+0.027}_{-0.016}$&0.036$^{+0.018}_{-0.011}$&0.025$^{+0.011}_{-0.007}$&0.018$^{+0.007}_{-0.006}$&0.013$^{+0.006}_{-0.004}$&0.011$^{+0.004}_{-0.003}$\\
&&&0.066&0.045&0.034&0.034&0.025&0.016\\
0.0&0.8&0.2&0.047$^{+0.022}_{-0.014}$&0.032$^{+0.014}_{-0.010}$&0.025$^{+0.012}_{-0.007}$&0.020$^{+0.010}_{-0.006}$&0.016$^{+0.007}_{-0.005}$&0.013$^{+0.005}_{-0.004}$\\
&&&0.049&0.032&0.033&0.032&0.021&0.015\\
0.2&0.6&0.2&0.056$^{+0.029}_{-0.018}$&0.036$^{+0.018}_{-0.012}$&0.027$^{+0.013}_{-0.008}$&0.021$^{+0.011}_{-0.006}$&0.016$^{+0.008}_{-0.005}$&0.013$^{+0.006}_{-0.004}$\\
&&&0.055&0.037&0.038&0.035&0.026&0.016\\
0.4&0.4&0.2&0.062$^{+0.032}_{-0.020}$&0.040$^{+0.018}_{-0.012}$&0.029$^{+0.014}_{-0.009}$&0.022$^{+0.010}_{-0.007}$&0.016$^{+0.008}_{-0.004}$&0.013$^{+0.006}_{-0.004}$\\
&&&0.071&0.044&0.039&0.041&0.030&0.018\\
0.6&0.2&0.2&0.068$^{+0.035}_{-0.022}$&0.043$^{+0.022}_{-0.013}$&0.031$^{+0.013}_{-0.011}$&0.022$^{+0.010}_{-0.006}$&0.016$^{+0.007}_{-0.005}$&0.012$^{+0.006}_{-0.003}$\\
&&&0.080&0.053&0.039&0.042&0.026&0.018\\
0.8&0.0&0.2&0.066$^{+0.033}_{-0.021}$&0.044$^{+0.021}_{-0.014}$&0.029$^{+0.014}_{-0.009}$&0.020$^{+0.009}_{-0.006}$&0.015$^{+0.006}_{-0.005}$&0.011$^{+0.006}_{-0.003}$\\
&&&0.083&0.051&0.041&0.038&0.026&0.017\\
0.0&0.6&0.4&0.061$^{+0.033}_{-0.019}$&0.039$^{+0.018}_{-0.012}$&0.030$^{+0.013}_{-0.010}$&0.021$^{+0.010}_{-0.006}$&0.017$^{+0.006}_{-0.005}$&0.012$^{+0.006}_{-0.004}$\\
&&&0.071&0.042&0.039&0.036&0.023&0.015\\
0.2&0.4&0.4&0.074$^{+0.041}_{-0.024}$&0.044$^{+0.023}_{-0.013}$&0.032$^{+0.015}_{-0.010}$&0.022$^{+0.012}_{-0.006}$&0.016$^{+0.008}_{-0.004}$&0.013$^{+0.005}_{-0.004}$\\
&&&0.079&0.053&0.048&0.040&0.024&0.016\\
0.4&0.2&0.4&0.082$^{+0.048}_{-0.026}$&0.050$^{+0.023}_{-0.016}$&0.034$^{+0.015}_{-0.011}$&0.023$^{+0.010}_{-0.007}$&0.017$^{+0.007}_{-0.005}$&0.012$^{+0.005}_{-0.003}$\\
&&&0.102&0.061&0.052&0.041&0.028&0.015\\
0.6&0.0&0.4&0.082$^{+0.043}_{-0.024}$&0.050$^{+0.025}_{-0.015}$&0.033$^{+0.013}_{-0.011}$&0.022$^{+0.009}_{-0.007}$&0.016$^{+0.007}_{-0.005}$&0.012$^{+0.005}_{-0.004}$\\
&&&0.110&0.063&0.053&0.039&0.025&0.016\\
0.0&0.4&0.6&0.082$^{+0.042}_{-0.026}$&0.048$^{+0.023}_{-0.014}$&0.033$^{+0.016}_{-0.010}$&0.022$^{+0.010}_{-0.006}$&0.016$^{+0.008}_{-0.005}$&0.011$^{+0.006}_{-0.003}$\\
&&&0.093&0.057&0.049&0.038&0.022&0.014\\
0.2&0.2&0.6&0.097$^{+0.055}_{-0.032}$&0.052$^{+0.026}_{-0.016}$&0.034$^{+0.016}_{-0.010}$&0.022$^{+0.011}_{-0.006}$&0.016$^{+0.006}_{-0.005}$&0.012$^{+0.005}_{-0.004}$\\
&&&0.120&0.064&0.050&0.036&0.023&0.015\\
0.4&0.0&0.6&0.102$^{+0.052}_{-0.033}$&0.053$^{+0.028}_{-0.017}$&0.034$^{+0.014}_{-0.010}$&0.022$^{+0.009}_{-0.007}$&0.015$^{+0.007}_{-0.004}$&0.011$^{+0.005}_{-0.003}$\\
&&&0.129&0.072&0.049&0.039&0.024&0.015\\
0.0&0.2&0.8&0.096$^{+0.046}_{-0.030}$&0.053$^{+0.025}_{-0.017}$&0.029$^{+0.013}_{-0.009}$&0.018$^{+0.008}_{-0.006}$&0.013$^{+0.007}_{-0.004}$&0.009$^{+0.005}_{-0.002}$\\
&&&0.159&0.064&0.031&0.017&0.009&0.011\\
0.2&0.0&0.8&0.104$^{+0.055}_{-0.035}$&0.054$^{+0.027}_{-0.016}$&0.029$^{+0.015}_{-0.009}$&0.018$^{+0.009}_{-0.006}$&0.013$^{+0.007}_{-0.004}$&0.011$^{+0.004}_{-0.004}$\\
&&&0.215&0.072&0.038&0.018&0.010&0.014\\
0.0&0.0&1.0&0.075$^{+0.037}_{-0.024}$&0.038$^{+0.020}_{-0.012}$&0.020$^{+0.010}_{-0.006}$&0.013$^{+0.007}_{-0.004}$&0.010$^{+0.004}_{-0.003}$&0.007$^{+0.004}_{-0.002}$\\
&&&0.129&0.041&0.023&0.011&0.007&0.008\\
\end{scotch}
\end{table*}

\begin{table*}[htbp]
\topcaption{Lower limits on the mass of the \cPQT quark at 95\% CL, for
  different combinations of \cPQT quark branching fractions. The $1\sigma$
  uncertainty range on the expected limits are given as well.}
\centering
\begin{scotch}{cccccc}
$\mathcal{B}(\tH)$ & $\mathcal{B}(\tZ)$ & $\mathcal{B}(\bW)$ &
Obs. limit & Exp. limit & Expected $1\sigma$\\
\hline
0.0 & 1.0 & 0.0 & 790 & 830 & [790,880] \\
0.2 & 0.8 & 0.0 & 780 & 820 & [780,870] \\
0.4 & 0.6 & 0.0 & 760 & 810 & [770,870] \\
0.6 & 0.4 & 0.0 & 760 & 820 & [770,870] \\
0.8 & 0.2 & 0.0 & 760 & 830 & [780,880] \\
1.0 & 0.0 & 0.0 & 770 & 840 & [780,890] \\
0.0 & 0.8 & 0.2 & 770 & 810 & [770,870] \\
0.2 & 0.6 & 0.2 & 760 & 800 & [760,870] \\
0.4 & 0.4 & 0.2 & 750 & 800 & [760,870] \\
0.6 & 0.2 & 0.2 & 750 & 800 & [760,870] \\
0.8 & 0.0 & 0.2 & 750 & 810 & [770,880] \\
0.0 & 0.6 & 0.4 & 760 & 800 & [760,870] \\
0.2 & 0.4 & 0.4 & 730 & 800 & [750,860] \\
0.4 & 0.2 & 0.4 & 720 & 790 & [740,860] \\
0.6 & 0.0 & 0.4 & 720 & 800 & [750,870] \\
0.0 & 0.4 & 0.6 & 740 & 800 & [750,860] \\
0.2 & 0.2 & 0.6 & 740 & 800 & [740,870] \\
0.4 & 0.0 & 0.6 & 730 & 800 & [750,870] \\
0.0 & 0.2 & 0.8 & 890 & 840 & [780,890] \\
0.2 & 0.0 & 0.8 & 870 & 840 & [770,890] \\
0.0 & 0.0 & 1.0 & 920 & 890 & [810,950] \\
\end{scotch}
\label{tab:limit_table}
\end{table*}


Depending on the assumed branching fractions, the expected limits
lie between 790 and 890\GeV, while the observed limits are
in a range between 720 and 920\GeV. In much of the triangular plane of
branching fractions these are the most stringent limits on \cPQT quark pair
production to date.

\section{Summary}
A search for pair production of vector-like \cPQT quarks of  charge $\twothird$ has been performed. In most models the hypothetical \cPQT quark has three decay modes: $ \cPQT\to \tH$, $ \cPQT \to \tZ$, and $ \cPQT \to \bW$.  The following five distinct topologies have been investigated: inclusive lepton events covering all possible decay modes,  single-lepton events  optimized to find  $ \cPQT \to \bW$ decays, all-hadronic events optimized either for $ \cPQT\to \tH$ or $ \cPQT\to \bW$ decays, and events containing a Higgs boson decaying to a pair of photons.

Data and SM background expectations are found to be in agreement. Upper limits on the production cross sections of vector-like \cPQT quarks are set. The expected 95\% CL lower mass limits are between 790 and 890\GeV depending on the branching fraction of the \cPQT quark. For a branching fraction of $\mathcal{B}(\tH)=100\%$  an expected (observed) limit of 840 (770)\GeV is found. For $\mathcal{B}(\tZ)=100\%$  the expected (observed) limit is 830 (790)\GeV and for $\mathcal{B}(\bW)=100\%$ the limit is 890 (920)\GeV. These are among the strongest limits on vector-like \cPQT quarks obtained to date.

\begin{acknowledgments}
\hyphenation{Bundes-ministerium Forschungs-gemeinschaft Forschungs-zentren} We congratulate our colleagues in the CERN accelerator departments for the excellent performance of the LHC and thank the technical and administrative staffs at CERN and at other CMS institutes for their contributions to the success of the CMS effort. In addition, we gratefully acknowledge the computing centers and personnel of the Worldwide LHC Computing Grid for delivering so effectively the computing infrastructure essential to our analyses. Finally, we acknowledge the enduring support for the construction and operation of the LHC and the CMS detector provided by the following funding agencies: the Austrian Federal Ministry of Science, Research and Economy and the Austrian Science Fund; the Belgian Fonds de la Recherche Scientifique, and Fonds voor Wetenschappelijk Onderzoek; the Brazilian Funding Agencies (CNPq, CAPES, FAPERJ, and FAPESP); the Bulgarian Ministry of Education and Science; CERN; the Chinese Academy of Sciences, Ministry of Science and Technology, and National Natural Science Foundation of China; the Colombian Funding Agency (COLCIENCIAS); the Croatian Ministry of Science, Education and Sport, and the Croatian Science Foundation; the Research Promotion Foundation, Cyprus; the Ministry of Education and Research, Estonian Research Council via IUT23-4 and IUT23-6 and European Regional Development Fund, Estonia; the Academy of Finland, Finnish Ministry of Education and Culture, and Helsinki Institute of Physics; the Institut National de Physique Nucl\'eaire et de Physique des Particules~/~CNRS, and Commissariat \`a l'\'Energie Atomique et aux \'Energies Alternatives~/~CEA, France; the Bundesministerium f\"ur Bildung und Forschung, Deutsche Forschungsgemeinschaft, and Helmholtz-Gemeinschaft Deutscher Forschungszentren, Germany; the General Secretariat for Research and Technology, Greece; the National Scientific Research Foundation, and National Innovation Office, Hungary; the Department of Atomic Energy and the Department of Science and Technology, India; the Institute for Studies in Theoretical Physics and Mathematics, Iran; the Science Foundation, Ireland; the Istituto Nazionale di Fisica Nucleare, Italy; the Ministry of Science, ICT and Future Planning, and National Research Foundation (NRF), Republic of Korea; the Lithuanian Academy of Sciences; the Ministry of Education, and University of Malaya (Malaysia); the Mexican Funding Agencies (CINVESTAV, CONACYT, SEP, and UASLP-FAI); the Ministry of Business, Innovation and Employment, New Zealand; the Pakistan Atomic Energy Commission; the Ministry of Science and Higher Education and the National Science Centre, Poland; the Funda\c{c}\~ao para a Ci\^encia e a Tecnologia, Portugal; JINR, Dubna; the Ministry of Education and Science of the Russian Federation, the Federal Agency of Atomic Energy of the Russian Federation, Russian Academy of Sciences, and the Russian Foundation for Basic Research; the Ministry of Education, Science and Technological Development of Serbia; the Secretar\'{\i}a de Estado de Investigaci\'on, Desarrollo e Innovaci\'on and Programa Consolider-Ingenio 2010, Spain; the Swiss Funding Agencies (ETH Board, ETH Zurich, PSI, SNF, UniZH, Canton Zurich, and SER); the Ministry of Science and Technology, Taipei; the Thailand Center of Excellence in Physics, the Institute for the Promotion of Teaching Science and Technology of Thailand, Special Task Force for Activating Research and the National Science and Technology Development Agency of Thailand; the Scientific and Technical Research Council of Turkey, and Turkish Atomic Energy Authority; the National Academy of Sciences of Ukraine, and State Fund for Fundamental Researches, Ukraine; the Science and Technology Facilities Council, UK; the US Department of Energy, and the US National Science Foundation.

Individuals have received support from the Marie-Curie programme and the European Research Council and EPLANET (European Union); the Leventis Foundation; the A. P. Sloan Foundation; the Alexander von Humboldt Foundation; the Belgian Federal Science Policy Office; the Fonds pour la Formation \`a la Recherche dans l'Industrie et dans l'Agriculture (FRIA-Belgium); the Agentschap voor Innovatie door Wetenschap en Technologie (IWT-Belgium); the Ministry of Education, Youth and Sports (MEYS) of the Czech Republic; the Council of Science and Industrial Research, India; the HOMING PLUS programme of the Foundation for Polish Science, cofinanced from European Union, Regional Development Fund; the Compagnia di San Paolo (Torino); the Consorzio per la Fisica (Trieste); MIUR project 20108T4XTM (Italy); the Thalis and Aristeia programmes cofinanced by EU-ESF and the Greek NSRF; the National Priorities Research Program by Qatar National Research Fund; the Rachadapisek Sompot Fund for Postdoctoral Fellowship, Chulalongkorn University (Thailand); and the Welch Foundation.
\end{acknowledgments}
\bibliography{auto_generated}

\cleardoublepage \appendix\section{The CMS Collaboration \label{app:collab}}\begin{sloppypar}\hyphenpenalty=5000\widowpenalty=500\clubpenalty=5000\textbf{Yerevan Physics Institute,  Yerevan,  Armenia}\\*[0pt]
V.~Khachatryan, A.M.~Sirunyan, A.~Tumasyan
\vskip\cmsinstskip
\textbf{Institut f\"{u}r Hochenergiephysik der OeAW,  Wien,  Austria}\\*[0pt]
W.~Adam, E.~Asilar, T.~Bergauer, J.~Brandstetter, E.~Brondolin, M.~Dragicevic, J.~Er\"{o}, M.~Flechl, M.~Friedl, R.~Fr\"{u}hwirth\cmsAuthorMark{1}, V.M.~Ghete, C.~Hartl, N.~H\"{o}rmann, J.~Hrubec, M.~Jeitler\cmsAuthorMark{1}, V.~Kn\"{u}nz, A.~K\"{o}nig, M.~Krammer\cmsAuthorMark{1}, I.~Kr\"{a}tschmer, D.~Liko, T.~Matsushita, I.~Mikulec, D.~Rabady\cmsAuthorMark{2}, B.~Rahbaran, H.~Rohringer, J.~Schieck\cmsAuthorMark{1}, R.~Sch\"{o}fbeck, J.~Strauss, W.~Treberer-Treberspurg, W.~Waltenberger, C.-E.~Wulz\cmsAuthorMark{1}
% ...
\vskip\cmsinstskip
% ...
1:~~Also at Vienna University of Technology, Vienna, Austria\\
2:~~Also at CERN, European Organization for Nuclear Research, Geneva, Switzerland\\
% ...

\end{sloppypar}
\end{document}